\documentclass[aps,prd,twocolumn,superscriptaddress,showpacs]{revtex4}
\usepackage[dvipdfmx]{graphicx}
\usepackage{here}
\usepackage{amsmath,amssymb,times}
\usepackage{color}
\usepackage{mathrsfs}

\newcommand{\bequ}{\begin{equation}}
\newcommand{\eequ}{\end{equation}}
\newcommand{\bea}{\begin{eqnarray}}
\newcommand{\eea}{\end{eqnarray}}

\def\gsim{~\,\makebox(1,1){$\stackrel{>}{\widetilde{}}$}\,~}
\def\lsim{~\,\makebox(1,1){$\stackrel{<}{\widetilde{}}$}\,~}

\DeclareSymbolFont{boldletters}{OML}{cmm} {b}{it}
\DeclareSymbolFontAlphabet{\mathbit}{boldletters}
\DeclareMathSymbol{\alpha}{\mathalpha}{letters}{"0B}
\DeclareMathSymbol{\beta}{\mathalpha}{letters}{"0C}
\DeclareMathSymbol{\gamma}{\mathalpha}{letters}{"0D}
\DeclareMathSymbol{\delta}{\mathalpha}{letters}{"0E}
\DeclareMathSymbol{\varepsilon}{\mathalpha}{letters}{"0F}
\DeclareMathSymbol{\zeta}{\mathalpha}{letters}{"10}
\DeclareMathSymbol{\eta}{\mathalpha}{letters}{"11}
\DeclareMathSymbol{\theta}{\mathalpha}{letters}{"12}
\DeclareMathSymbol{\iota}{\mathalpha}{letters}{"13}
\DeclareMathSymbol{\kappa}{\mathalpha}{letters}{"14}
\DeclareMathSymbol{\lambda}{\mathalpha}{letters}{"15}
\DeclareMathSymbol{\mu}{\mathalpha}{letters}{"16}
\DeclareMathSymbol{\nu}{\mathalpha}{letters}{"17}
\DeclareMathSymbol{\xi}{\mathalpha}{letters}{"18}
\DeclareMathSymbol{\pi}{\mathalpha}{letters}{"19}
\DeclareMathSymbol{\rho}{\mathalpha}{letters}{"1A}
\DeclareMathSymbol{\sigma}{\mathalpha}{letters}{"1B}
\DeclareMathSymbol{\tau}{\mathalpha}{letters}{"1C}
\DeclareMathSymbol{\upsilon}{\mathalpha}{letters}{"1D}
\DeclareMathSymbol{\phi}{\mathalpha}{letters}{"1E}
\DeclareMathSymbol{\chi}{\mathalpha}{letters}{"1F}
\DeclareMathSymbol{\psi}{\mathalpha}{letters}{"20}
\DeclareMathSymbol{\omega}{\mathalpha}{letters}{"21}
\DeclareMathSymbol{\varepsilon}{\mathalpha}{letters}{"22}
\DeclareMathSymbol{\vartheta}{\mathalpha}{letters}{"23}
\DeclareMathSymbol{\varpi}{\mathalpha}{letters}{"24}
\DeclareMathSymbol{\varrho}{\mathalpha}{letters}{"25}
\DeclareMathSymbol{\varsigma}{\mathalpha}{letters}{"26}
\DeclareMathSymbol{\varphi}{\mathalpha}{letters}{"27}
\DeclareMathSymbol{\Gamma}{\mathalpha}{letters}{"00}
\DeclareMathSymbol{\Delta}{\mathalpha}{letters}{"01}
\DeclareMathSymbol{\Theta}{\mathalpha}{letters}{"02}
\DeclareMathSymbol{\Lambda}{\mathalpha}{letters}{"03}
\DeclareMathSymbol{\Xi}{\mathalpha}{letters}{"04}
\DeclareMathSymbol{\Pi}{\mathalpha}{letters}{"05}
\DeclareMathSymbol{\Sigma}{\mathalpha}{letters}{"06}
\DeclareMathSymbol{\Upsilon}{\mathalpha}{letters}{"07}
\DeclareMathSymbol{\Phi}{\mathalpha}{letters}{"08}
\DeclareMathSymbol{\Psi}{\mathalpha}{letters}{"09}
\DeclareMathSymbol{\Omega}{\mathalpha}{letters}{"0A}

\begin{document}
\title{
A hadron-quark hybrid model reliable for the EoS in $\mu_{B} \leq 400$~MeV
}

\author{Akihisa Miyahara}
\affiliation{Observation Division, Chubu aviation weather service center, Japan Meteorological Agency, Tokoname 479-0881, Japan}

\author{Masahiro Ishii}
\affiliation{Department of Physics, Graduate School of Sciences, Kyushu University,
             Fukuoka 819-0395, Japan}

\author{Hiroaki Kouno}
\affiliation{Department of Physics, Saga University,
             Saga 840-8502, Japan}  

\author{Masanobu Yahiro}
\email[]{orion093g@gmail.com}
\affiliation{Department of Physics, Graduate School of Sciences, Kyushu University,
             Fukuoka 819-0395, Japan}             
\date{\today}

\begin{abstract}
We present a simple version of hadron-quark hybrid (HQH) model in the $\mu_B$--$T$ plain, 
where $T$ is temperature and  $\mu_{B}$ is the baryon-number chemical potential.
The model is composed of the independent-quark model
for quark-gluon states and an improved version 
of excluded-volume hadron resonance gas (EV-HRG) model for 
hadronic states. 
In the improved version of EV-HRG, the pressure  has 
charge conjugation and is obtained by a simple analytic form. 
The switching function from hadron states to  quark-gluon states 
in the present model  has no chemical potential dependence. 
The simple HQH model  
is successful in reproducing LQCD results on the transition region of chiral crossover and 
the EoS in $\mu_{B} \leq 400$~MeV. We then predict the chiral-crossover  region in   $400 \leq \mu_{B} \leq 800$~MeV. 
We also predict a transition line derived from isentropic trajectories in $0 \leq \mu_{B} \leq 800$~MeV and 
find that  the effect of strangeness neutrality is small there. 
\end{abstract} 
\maketitle

\section{Introduction}
\label{Introduction}

 {\it LQCD:}
  The state-of-art 2+1-flavor lattice QCD (LQCD) simulation of Ref.~\cite{YAoki_crossover}
 showed that the  transition is ``crossover'' at finite temperature ($T$) and zero baryon chemical
 potential ($\mu_{B} = 0$), where the continuum and thermodynamic limits
 were carefully taken. In general, the crossover
 nature means that the transition temperature depends on the
 choice of observables. In fact, observable-dependent transition
 temperatures $T_c^{(O)}(\mu_{B})$ have been discussed in
 LQCD simulations for zero and small 
 $\mu_{B}$; actually,  
the renormalized chiral condensate $O=\Delta_{l,s}(T,\mu_B)$, 
the Polykov loop $O=\Phi(T,\mu_B)$, the energy density $O=\varepsilon(T,\mu_B)$ and 
the trace anomaly $O=I(T,\mu_B)$ are taken in Refs.~\cite{Fodor:2004nz,
YAoki_Tc,Aoki:2009sc,Borsanyi:2010bp,Endrodi:2011gv,Borsanyi:2012cr,Bellwied:2015rza,Bazavov:2017dus}. 
In Ref.~\cite{Bazavov:2017dus}, the LQCD data disfavors 
the existence of critical endpoint (CEP) in 
$\mu_{B}/T \leq 2$ and $T/T_c^{(\Delta_{l,s})} (\mu_{B}= 0) >  0.9$. 
The equation of state (EoS) is  important particularly for relativistic nuclear collisions and neutron stars. 
The location of transition region is essential to determine the EoS. 
For these reasons,  a lot of LQCD data have been accumulated~\cite{YAoki_crossover,Fodor:2004nz,
YAoki_Tc,Aoki:2009sc,Borsanyi:2010bp,Endrodi:2011gv,Bellwied:2015rza,Borsanyi:2012cr,Borsanyi_sus_plot,Borsanyi:2013bia,Bazavov:2014pvz,Borsanyi:2016ksw,Bazavov:2017dus}.

 {\it Effective models:}
As a complementary approach to LQCD simulations, 
we can consider effective models such as the quark-meson model~\cite{Jungnickel:1995fp} and 
the Polyakov-loop extended Nambu--Jona-Lasinio (PNJL) model~\cite{MO,Dumitru,Fukushima,Sakai}. 
The model approach is useful for the prediction of the transition lines, 
the presence or absence of the CEP  and the EoS. 
The hadron resonance gas (HRG) model is a simple model for
hadronic matter and remarkably reproduces LQCD data on the EoS in $T \lsim 1.3
T_c^{(\Delta_{l,s})} (\mu_{B}= 0)$~\cite{Borsanyi:2013bia}.

As a simplified version of the PNJL model~\cite{MO,Dumitru,Fukushima,Sakai},  
the independent quark (IQ) model
reproduces $T$ dependence of the Polyakov loop calculated with
2+1-flavor LQCD simulations for 
$\mu_{B}= 0$~\cite{Miyahara:2016din,Miyahara:2017eam}, although the PNJL model does not.  
The IQ model treats the coupling between the 
quark field and the homogeneous classical gauge field,
 but not the couplings between quarks.

\begin{figure}[H]
\centering
\vspace{0cm}
\includegraphics[width=0.3\textwidth]{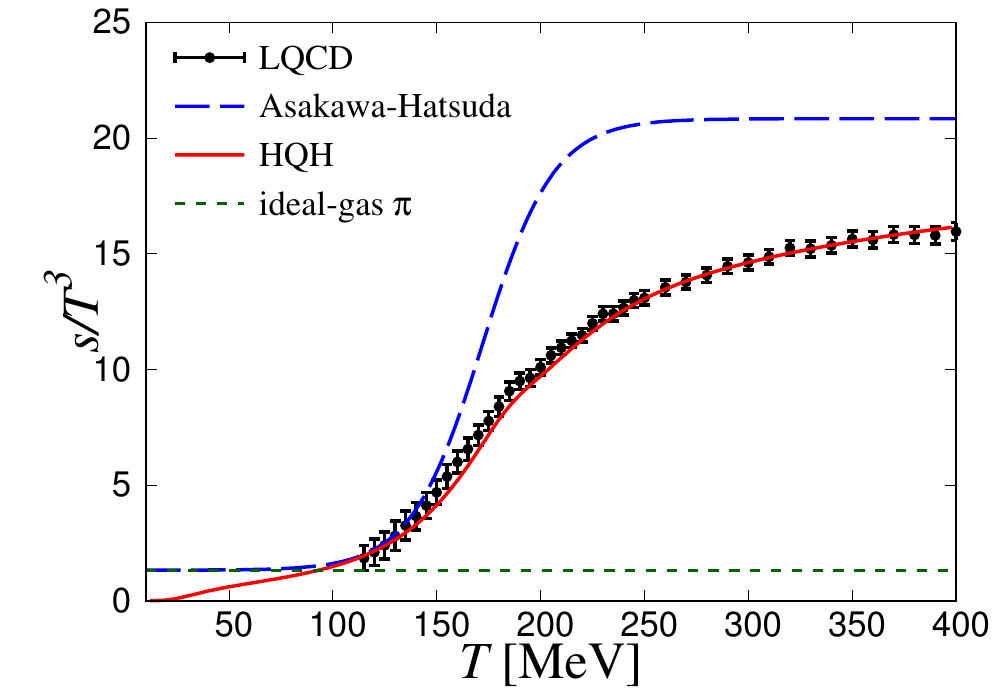}
\vspace{-10pt}
\caption{$T$ dependence of $s/T^3$ at $\mu_{B}=0$~MeV. 
The dashed line denotes the  $s/T^3$ of Ref.~\cite{Hatsuda_hybrid model} in which 
$s_{\rm Q}(T)/T^3 \equiv 190(\pi^2/90)$ for three-flavor free quark-gluon gas and 
$T_c=172$~MeV, where $T_c$ is the chiral pseudo-critical temperature at  $\mu_{B}=0$~MeV. 
The solid line stands for the result of  Ref.~\cite{Miyahara:2017eam}. 
The dotted line is $s_{\rm H}(T)/T^3 \equiv 12(\pi^2/90)$ of massless pion gas. 
LQCD data for 2+1 flavor are taken from Ref.~\cite{Borsanyi:2012cr}.
}
\label{fig_Eos-AH}
\end{figure}

{\it Hadron-quark hybrid (HQH) model:}
Asakawa and Hatsuda proposed the HQH model for 
$\mu_B=0$ in order to describe the coexistence of quarks and hadrons~\cite{Hatsuda_hybrid model}. 
The total entropy $s(T,\mu_B)$ of the model is      
$
s(T)=f_{\rm H}(T) s_{\rm H}(T) 
 + \left[1-f_{\rm H}(T) \right] 
s_{\rm Q}(T)  ,   
$
where $s_{\rm H}(T) \equiv 12(\pi^2/90)T^3$ 
and $s_{\rm Q}(T) \equiv 148(\pi^2/90)T^3$ are the entropy densities of massless
free gas with two flavors in the hadronic phase (pion gas) and in the quark-gluon phase, respectively.
The weight function $f_{\rm H}(T)$ means the occupancy of
hadronic matter in the total entropy, and assumed a simple function satisfying the condition 
$0\le f_{\rm H}\le 1$. 
As shown in Fig.~\ref{fig_Eos-AH}, their $s(T)$  (dashed line) does not  reproduce $s_{\rm LQCD}$, 
where the $f_{\rm H}(T)$ has a width parameter $\Gamma/T_c$ and the value 0.2 has been determined to reproduce 
the low $T$ part of  $s_{\rm LQCD}(T)$. 
In addition,  their $s(T)/T^3$ does not vanish at $T=0$, but the HRG does. 
Therefore, we should take the HRG as $s_{\rm H}(T)$ and IQ model  as  $s_{\rm Q}(T)$.

In our previous papers~\cite{Miyahara:2016din,Miyahara:2017eam}, we improved the HQH model 
of Ref.~\cite{Hatsuda_hybrid model} for 
finite $\mu_B$, taking the HRG model for the hadronic part and the independent quark (IQ) model  
for the quark-gluon part. The total entropy $s(T,\mu_B)$ reads    
\begin{eqnarray}
s(T,\mu_B)&=&f_{\rm H}(T,\mu_B) s_{\rm H}(T,\mu_B) 
\nonumber \\
&~& + \left[1-f_{\rm H}(T,\mu_B) \right] 
s_{\rm Q}(T,\mu_B)  .   
\label{s_hybrid_def}
\end{eqnarray}
The result (solid line) of Ref.~\cite{Miyahara:2017eam} reproduces LQCD data~\cite{Borsanyi:2012cr}, 
as shown in Fig.~\ref{fig_Eos-AH}.

Another type of HQH model was proposed in Refs~\cite{Albright:2014gva,Albright:2015uua}. The HQH model  
consider 
the pressure instead of the entropy. As an advantage of our approach, $s_{\rm LQCD}$ automatically 
satisfies the thermodynamic inequality and the Nernst's theorem~\cite{Landau-Lifshitz}, 
\bea
\left. \frac{\partial s(T,\mu_B)}{\partial T}\right|_{\mu_B=0 }> 0,
~~~
\left. s(T,\mu_B) \right|_{T=\mu_B=0} =0.
\label{Nernst's theorem}
\eea 
In our previous papers~\cite{Miyahara:2016din,Miyahara:2017eam}, the $f_{\rm H}(T,\mu_B)$ 
was determined from  
LQCD data on $s_{\rm LQCD}$ and the second-order susceptibilities at $\mu_B=0$. 
For this reason, the approach is applicable only for small $\mu_B$. 
We could not show the chiral-transition line, since $\Delta_{\rm l,s}$ becomes negative 
in $T \gsim 170$~MeV.

In the HRG model, the interactions between baryons (anti-baryon) are
neglected, but it should be taken into account for $\mu_{B}$
dependence of thermodynamic quantities. 
A simple way of treating volume-exclusion effects (repulsive force)~\cite{Kouno-VE}
was suggested in Refs.~\cite{Vovchenko:2014pka,Vovchenko:2015cbk}. 
This model is called ``excluded-volume HRG 
(EV-HRG) model''. Furthermore, a method of treating an attractive force 
in addition to the repulsive force was proposed 
in Ref.~\cite{Vovchenko:2016rkn}. 
The volume-exclusion effects are included 
by fitting the volume parameter $b=4 \cdot 4\pi r^3/3$~\cite{Landau-Lifshitz} to 
either LQCD data or the core radius $r$ 
of nucleon-nucleon force~\cite{Vovchenko:2014pka,Vovchenko:2015cbk}. 
In the framework of 
Refs.~\cite{Vovchenko:2014pka,Vovchenko:2015cbk,Vovchenko:2016rkn}, 
the interaction between baryon and anti-baryon and the 
radius of meson are neglected.

{\it Our aim:} 
In this paper, we improve  the HQH model 
of Ref.~\cite{Miyahara:2017eam}, taking the EV-HRG model 
for the hadron piece and the simple IQ model for the quark-gluon piece. 
The EV-HRG model taken yields the pressure as a simple analytic function 
and guarantees that the pressure is $\mu_B$ even. 
We refer to the present version of  HQH model as ``simple  HQH (sHQH) model''.

The present sHQH model have only six parameters, i.~e.,  one parameter $r$ in the EV-HRG model and 
five parameters in the IQ model. In the IQ model, the parameters are fitted to $s_{\rm LQCD}$
in $400 < T < 800$~MeV and $\mu_{B}=0$~\cite{Miyahara:2017eam}. 
In our EV-HRG model,  as a value of $r$, we take the hard-core radius $r=0.34$~fm 
of the Hamada-Johnston nucleon-nucleon  interaction~\cite{Hamada-Johnston}, since 
the other nuclear forces do not have the hard core. 
We have also supposed that the hard core universally emerges in the other baryon-baryon interactions between hyperons or excited baryons, and their core radii are assumed to be the same as that of nucleon. 
We then determined the switching function $f_{\rm H}$ from $s_{\rm LQCD}$ at $\mu_{B}=0$. 
The sHQH model  with the  $f_{\rm H}(T,0)$ reproduces  LQCD data on 
the Polyakov loop at zero chemical potential and the EoS in finite $\mu_{B}$ up to 400~MeV. 
The present sHQH model thus has no $\mu_{B}$ in  $f_{\rm H}$; namely, 
$\mu_{B}$ dependence of physical quantities come from the EV-HRG and the IQ model. 
We thus succeed in simplifying the HQH model by taking $r=0.34$~fm.

The $\Delta_{\rm l,s}$ signals  the chiral transition. 
The  $\Delta_{\rm l,s}$ calculated with the HRG model becomes 
negative in $T \gsim 170$~MeV~\cite{Borsanyi:2010bp}, whereas 
the corresponding LQCD result is positive. 
The present model have this problem. 
We circumvent this problem in the following way.

As an interesting result of LQCD simulations in Ref.~\cite{Borsanyi:2010bp}, 
the chirla-crossover region determined from $d \Delta_{\rm l,s}/dT$ 
agrees with that from  $d \varepsilon/dT$ at $\mu_B=0$. 
In LQCD simulations of Ref.~\cite{Borsanyi:2012cr}, furthermore, the transition region is obtained 
by $d \varepsilon/dT$ for finite $\mu_B$. 
Therefore, we use  the peak and the half-value width of  $d \varepsilon/dT$ 
as  a transition region in $\mu_B$--$T$ plane.  
We show that the transition region determined from  $\varepsilon$ agrees with the chiral-transition region 
calculated with LQCD simulations~\cite{Bellwied:2015rza}.  

As mentioned above, the present sHQH model well reproduces LQCD data on the EoS and the chiral-crossover region in  $0 \leq \mu_B \leq 400$~MeV. We can then predict the transition region of chiral crossover  
in  $400 \leq \mu_B \leq 800$~MeV. 
LQCD data will become available for $\mu_B=400 \sim 800$~MeV by development of LQCD simulations 
such as the complex Langevin method~\cite{Aarts1,Aarts2,Sexty,Aarts3}.

Finally, we present  a transition line derived from isentropic trajectories in $0 \leq \mu_{B} \leq 800$~MeV. 
When we calculate the isentropic trajectories, we switch on and off the strangeness neutrality. 
We find that the effect is small there. 
For this reason, we do not 
consider the strangeness neutrality for the chiral-crossover  region and the EoS.

This paper is organized as follows. 
In Sec.~\ref{model}, we show the model building. 
Numerical results are shown in Sec~\ref{results}. 
Section~\ref{summary} is devoted to a summary.

\vspace{-0pt}
\section{Model building}
\label{model}

We present a simple version of HQH model. 
The model is composed of an improved version 
of EV-HRG model for hadronic states and  
the IQ model for quark-gluon states.

For the 2+1 flavor system, we can consider the chemical potentials of u, d, s quarks by $\mu_{\rm u}, \mu_{\rm d}$, $\mu_{\rm s}$, respectively. These potentials are related to the baryon-number ($B$) chemical potential $\mu_{B}$, the isospin ($I$) chemical potential $\mu_{I}$ and the hypercharge ($Y$) chemical potential $\mu_{Y}$ as 
\begin{eqnarray}
\begin{array}{l}
\mu_{B} = \mu_{\rm u} + \mu_{\rm d} + \mu_{\rm s},\\
\mu_{I} = \mu_{\rm u} - \mu_{\rm d},\\
\mu_{Y} = \frac{1}{2}(\mu_{\rm u} + \mu_{\rm d} - 2\mu_{\rm s}). 
\end{array}
\label{chemical potential 2+1}
\end{eqnarray}
As for $\mu_{I}$ and $\mu_{Y}$, 
the right-hand side of Eq.~\eqref{chemical potential 2+1} 
comes from the diagonal elements of the matrix representation of 
Cartan algebra in $SU(3)$ group: 
$\mu_{I} = (1,-1,0)(\mu_{\rm u},\mu_{\rm d},\mu_{\rm s})^{\rm t}$ and 
$\mu_{Y} = (1/2)(1,1,-2)(\mu_{\rm u},\mu_{\rm d},\mu_{\rm s})^{\rm t}$. 
Equation \eqref{chemical potential 2+1} yields 
\begin{eqnarray}
\renewcommand{\arraystretch}{1.2}
\begin{array}{l}
\mu_{\rm u} = \frac{1}{3}\mu_{B} + \frac{1}{2}\mu_{I} + \frac{1}{3}\mu_{Y},\\
\mu_{\rm d} = \frac{1}{3}\mu_{B} - \frac{1}{2}\mu_{I} + \frac{1}{3}\mu_{Y},\\
\mu_{\rm s} = \frac{1}{3}\mu_{B} - \frac{2}{3}\mu_{Y}.
\end{array}
\label{2+1chemical potential}
\end{eqnarray}

\subsection{HRG model}
\label{HRG model}
For later convenience, we start with the HRG model.  
In the model, the pressure $P_{\rm H}$ is 
divided into the baryon (B) part  $P_{\rm B}$, the anti-baryon (aB) 
part $P_{\rm aB}$ and the meson (M) part $P_{\rm M}$:
\begin{eqnarray}
P_{\rm H}\equiv
{P_{\rm B}}+{P_{\rm aB}}+{P_{\rm M}}
\label{Pressure_HRG}
\end{eqnarray}
with
\begin{eqnarray}
P_{\rm B} = \sum_{i \in \rm B}d_{i}T\int  \log(1+ e^{-(E_{{\rm B},i} - \mu_{{\rm B},i})/T}),  
\label{Pressure_B}
\end{eqnarray}
\begin{eqnarray}
P_{\rm aB} = \sum_{i \in \rm aB}d_{i}T\int  \log(1+ e^{-(E_{{\rm B},i} + \mu_{{\rm B},i})/T}) ,
\label{Pressure_aB}
\end{eqnarray}
\begin{eqnarray}
P_{\rm M} &=& -\sum_{j \in \rm Meson}d_{j}T\int  \big\{ \log(1- e^{-(E_{\rm M,j}-\mu_{{\rm M},j})/T})\nonumber\\
&&+\log(1- e^{-(E_{{\rm M},j}+\mu_{{\rm M},j})/T})\big\}
\label{Pressure_M}
\end{eqnarray}
for $E_{{\rm B},i}=\sqrt{{\bf p}^2+{m_{{\rm B},i}}^2}$ and $E_{{\rm M},j}=\sqrt{{\bf p}^2+{m_{{\rm M},j}}^2}$, 
where $m_{{\rm B},i}$ ($m_{{\rm M},j}$) and $\mu_{{\rm B},i}$ ($\mu_{{\rm M},j}$) is the mass and the chemical potential of the $i$-th baryon 
($j$-th meson), respectively. Here we have used 
the shorthand notation 
\bea
\int \equiv  \int \frac{d^3{\bf p}}{(2\pi)^3}. 
\eea
for the  integration over 3d-momentum ${\bf p}$. 
In Eq. \eqref{Pressure_HRG}, all the hadrons listed in 
the Particle Data Table \cite{PDG} are taken.

\subsection{Improved version of EV-HRG}
\label{Improved version of EV-HRG}

We first explain the EV-HRG model of Refs.~\cite{Vovchenko:2016rkn,Vovchenko:2015cbk,Vovchenko:2014pka}.  
The  pressure  $P_{\rm EV;H}$ is obtained by 
\bea
{\rm P_{\rm EV;H}}={P_{\rm EV;B}}+{P_{\rm EV;aB}}+{P_{\rm M}} 
\label{EQ:P_EV-HRG}
\eea
with 
\begin{eqnarray}
{P_{\rm{EV;B}}} = \sum_{i \in \rm B}d_{i}T\int  \log(1+ e^{-(E_{{\rm B},i} - \mu_{{\rm EV:B},i})/T}),~~~  
\label{EQ:P_EV-B}
\end{eqnarray}
\begin{eqnarray}
{P_{\rm{EV;aB}}} = \sum_{i \in \rm aB}d_{i}T\int  \log(1+ e^{-(E_{{\rm B},i} + \mu_{{\rm EV:aB},i})/T}) .~~~
\label{EQ:P_EV-aB}
\end{eqnarray}
Here the effective baryon and anti-baryon chemical potentials, 
$\mu_{{\rm EV:B},i}$ and $\mu_{{\rm EV:aB},i}$,
 are defined by  
\bea
\mu_{{\rm EV:B},i}/T&=&\mu_{{B},i}/T-{\bar b} P_{\rm EV;B}/T^4,
\label{effective-mu-1}
\\
\mu_{{\rm EV:aB},i}/T&=&\mu_{{B},i}/T-{\bar b} P_{\rm EV;aB}/T^4,   
\label{effective-mu-2}
\eea
where ${\bar b}=bT^3$ for a positive volume parameter $b$. 
It is not easy to obtain 
$P_{\rm{EV;B}}$ and $P_{\rm{EV;aB}}$, since  $\mu_{{\rm EV;B},i}$ 
($\mu_{{\rm EV;aB},i}$) includes $P_{\rm{EV;B}}$ ($P_{\rm{EV;aB}}$) and
self-consistent calculation is necessary. 
Actually, $P_{\rm{EV;B}}$ and $P_{\rm{EV;aB}}$ are  obtained 
by solving Eqs. \eqref{EQ:P_EV-B} and \eqref{EQ:P_EV-aB} numerically.  
 
In QCD, the pressure is  charge-conjugation even ($\mu_B$ even). 
Hence the $P_{\rm EV;H}$ should be $\mu_B$ even,  
because it is a model of explaining QCD in $T < T_c$. 
Since $\mu_{{\rm EV:B},i}$ includes a 
$\mu_B$-odd term $\mu_{{B},i}/T$ and a $\mu_B$-even ${\bar b} P_{\rm EV;B}/T^4$, 
the resulting $P_{\rm EV;H}$  is not $\mu_B$ even. It is not natural.

We then redefine the ${P_{\rm{EV;B}}}$ and ${P_{\rm{EV;aB}}}$ so that the $P_{\rm EV;H}$ can be $\mu_B$-even. The redefined ${P_{\rm{EV;B}}}$ and ${P_{\rm{EV;aB}}}$ are denoted by 
${P_{\rm{inv;B}}}$ and ${P_{\rm{inv;aB}}}$, respectively: Namely,   
\begin{eqnarray}
{P_{\rm{inv;B}}} = \sum_{i \in \rm B}d_{i}T\int   \log(1+ e^{-(E_{{\rm B},i} - \mu_{{\rm inv;B},i})/T}),~~~
\label{EQ:P_mod-B}
\end{eqnarray}
\begin{eqnarray}
{P_{\rm{inv;aB}}} = \sum_{i \in \rm aB}d_{i}T\int  \log(1+ e^{-(E_{{\rm B},i} + \mu_{{\rm inv;aB},i})/T}) .~~~
\label{EQ:P_mod-aB}
\end{eqnarray}
with 
\bea
\mu_{{\rm inv:B},i}/T&=&\mu_{{B},i}/T-{\bar b} P_{\rm inv;B}/T^4,
\label{M-effective-mu-1}
\\
\mu_{{\rm inv:aB},i}/T&=&\mu_{{B},i}/T+{\bar b} P_{\rm inv;aB}/T^4,   
\label{M-effective-mu-2}
\eea
The sum of ${P_{\rm{inv;B}}}$ and ${P_{\rm{inv;aB}}}$ 
are $\mu_B$ even, since the sum is invariant under $\mu_B \rightarrow -\mu_B$. 
For this reason, we take Eqs.~\eqref{EQ:P_mod-B}--\eqref{M-effective-mu-2}. 
These equations  show that $P_{\rm{inv;B}} \geq P_{\rm{inv;aB}}$.

The $P_{\rm inv:B}$ and $P_{\rm inv:aB}$ can be rewritten into 
\bea
\frac{{P_{\rm{inv:B}}}}{T^4} &=& 
\sum_{i \in \rm B} A_i
 \sum_{\ell=1}^{\infty} \frac{(-1)^{\ell+1}}{ \ell^{2}} 
\nonumber \\
&\times &
K_2\Big({\ell {m_{i}} \over T}\Big)
\exp \Big({ {\ell \mu_{{\rm inv;B},i} \over T}}\Big),~~~~~~~
\label{EQ:P_mod:B-2} 
\eea
\begin{eqnarray}
\frac{{P_{\rm{inv:aB}}}}{T^4} &=& 
\sum_{i \in \rm aB} A_i
 \sum_{\ell=1}^{\infty} \frac{(-1)^{\ell+1}}{ \ell^{2}} 
\nonumber \\
&\times &
K_2\Big({\ell {m_{i}} \over T}\Big)
\exp \Big({- {\ell \mu_{{\rm inv;aB},i} \over T}}\Big)~~~~~~~
\label{EQ:P_mod:aB-2}
\end{eqnarray}
for 
\bea 
A_i \equiv \frac{d_i}{2{\pi}^2}\Big(\frac{m_{i}}{T}\Big)^2 .
\eea

LQCD data on the EoS are 
available for $T \le 400$~MeV and 
$\mu_{B} \le 400$~MeV~\cite{Borsanyi:2010bp,Borsanyi:2012cr}. 
We then consider this region.  
We consider $P_{\rm{B}}$, because of $P_{\rm{inv;B}} \geq P_{\rm{inv;aB}}$. 
The $\ell$ convergence of Eq.~\eqref{EQ:P_mod:B-2} becomes worse 
as $|(\mu_{B}-m_{i})/T|$ becomes larger; note that 
$K_2(x)$ is proportional to $\exp(-x)$ for large $x$ and 
$\mu_{B}-m_{i}$ is negative. 
Therefore, the convergence is worst for the smallest case 
$(939-400)/400$  where $T=\mu_{B}=400$~MeV and $m_N=939$~MeV. 
Taking the $\ell=1$ term only is a 3~\% error in 
Eqs.~\eqref{EQ:P_mod:B-2}. 
In actual calculations,  nucleon contribution in $P_{\rm{B}}$ is only 3~\%, 
so that  taking the $\ell=1$ term only corresponds to 0.1\% error. 
We can identify $P_{\rm{B}}$ with its $\ell=1$ term and  
$P_{\rm{aB}}$ with its $\ell=1$ one. This approximation is called 
``$\ell=1$ identification'' in this paper

Using the $\ell=1$ identification, we can rewrite $P_{\rm{inv:B}}$ as   
\bea
\frac{{P_{\rm{inv:B}}}}{T^4}
\simeq  \sum_{i \in \rm B} A_i 
K_2\Big({{m_{i}} \over T}\Big)
\exp \Big({ { \mu_{{\rm inv;B},i} \over T}}\Big),~~~
\label{EQ:P_mod-3} 
\eea
Multiplying both the sides of Eq.~\eqref{EQ:P_mod-3} by 
${\bar b}\exp ({\bar b}P_{{\rm inv;B}}/T^4)$ and using 
the $\ell=1$ identification, one can obtain

\bea
&&{\bar b}\frac{{P_{\rm{inv:B}}}}{T^4}
\exp \Big({\bar b}\frac{P_{\rm inv;B}}{T^4} \Big)
\nonumber \\
&=& 
{\bar b} \sum_{i \in \rm B} A_i K_2\Big({{m_{i}} \over T}\Big)
\exp \Big({ { \mu_{B,i} \over T}}\Big)
\nonumber \\
&=&{\bar b}\sum_{i \in \rm B} A_i
 \sum_{\ell=1}^{\infty} \frac{(-1)^{\ell+1}}{ \ell^{2}} 
K_2\Big({\ell {m_{i}} \over T}\Big)
\exp \Big({ {\ell \mu_{B,i} \over T}}\Big)
\nonumber \\
&=&{\bar b}\frac{{P_{\rm{inv;B}}( \mu_{B})  }}{T^4},~~~{\rm for}~~\mu_{B}=\mu_{B,i}.
\label{EQ:P_mod-derivation} 
\eea
Noting that the Lambert $W(z)$ function is the inverse function of $W e^W=z$, 
one can get $P_{\rm{inv:B}}$ as a simple analytic function: Namely,  
\bea
\frac{{P_{\rm{inv;B}}}}{T^4}=\frac{W({\bar b}P_{\rm{inv;B}}( \mu_{B}) /T^4)}{\bar b}.  
\label{EQ:PB_mod-ell=1}
\eea
In the limit of ${\bar b}=0$, the $P_{\rm{inv:B}}$ tends to  
$P_{\rm{B}}$, because of $W(z)  \rightarrow z$.  
Parallel discussion is possible for anti-baryon. The result is 
\bea
\frac{{P_{\rm{inv;aB}}}}{T^4}=\frac{W({\bar b}P_{\rm{aB}}( \mu_{B}) /T^4)}{\bar b}.  
\label{EQ:PaB_mod-ell=1}
\eea
Hence the hadronic pressure becomes 
\bea
{\rm P_{\rm inv;H}}={P_{\rm inv;B}}+{P_{\rm inv;aB}}+{P_{\rm M}} 
\label{EQ:P_EV-HRG-2}
\eea
with Eqs. \eqref{EQ:PB_mod-ell=1} and \eqref{EQ:PaB_mod-ell=1}. 
The entropy density $s_{\rm inv:H}$ is obtained from 
$P_{\rm {\rm inv:H}}$ as
\begin{eqnarray}
s_{\rm inv:H}&=&{\partial P_{\rm inv:H}\over{\partial T}}. 
\label{s_Q}
\end{eqnarray}
This improved version of EV-HRG model is referred to as 
``improved EV-HRG model'', but the difference between improved EV-HRG model and original EV-HRG model 
is not large for the pressure.

\begin{figure}[h]
\centering
\vspace{0cm}
\includegraphics[width=0.4\textwidth]{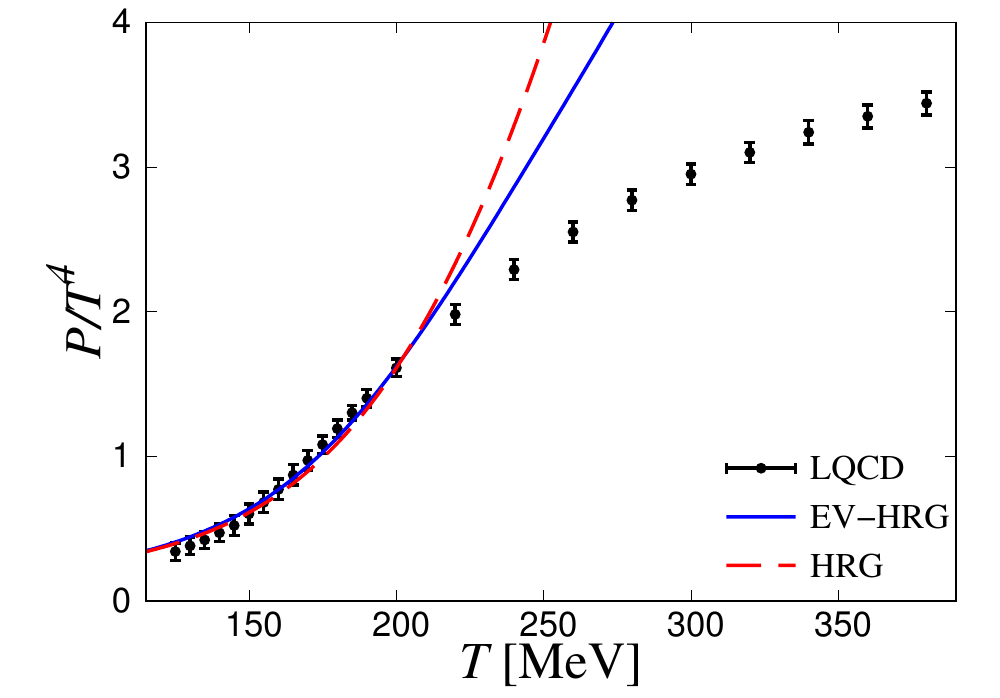}
\includegraphics[width=0.4\textwidth]{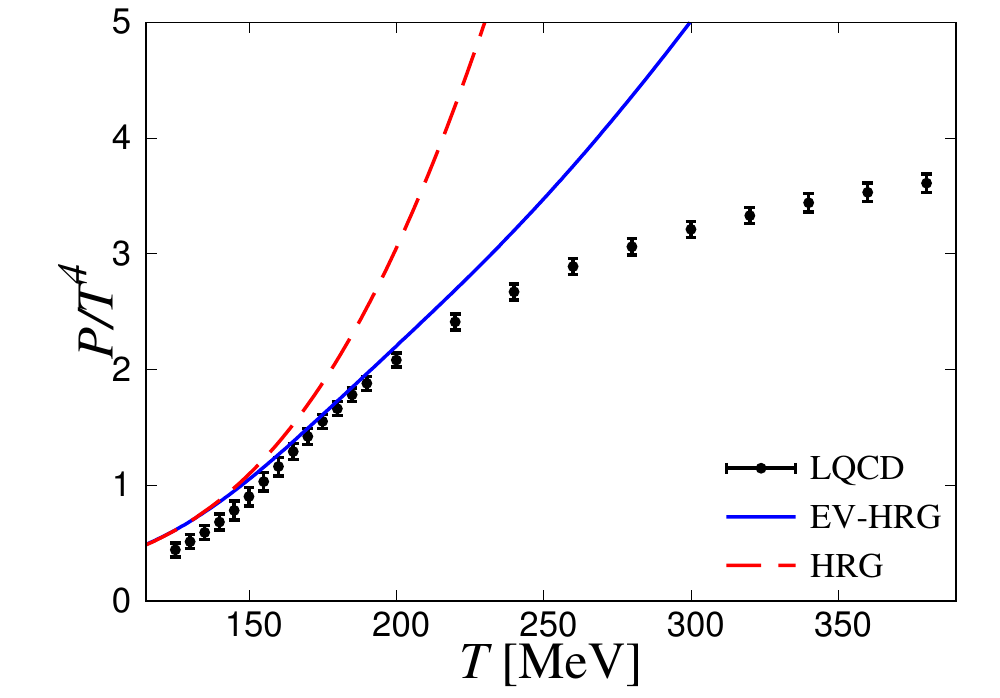}
\vspace{-10pt}
\caption{
$T$ dependence of pressure $P$ at $\mu_{B}=0$~MeV (upper panel) and 
$\mu_{B}=400$~MeV (lower panel).  
The solid and dashed lines stand for the results of improved EV-HRG model and 
HRG model, respectively. 
LQCD data are taken from Ref.~\cite{Borsanyi:2016ksw}. 
}
\label{fig_P-ev}
\end{figure}

Figure~\ref{fig_P-ev} shows $T$ dependence of the total pressure $P$ for $\mu_{B}=0, 400$~MeV. 
The results of improved EV-HRG and HRG models are compared 
with LQCD ones~\cite{Borsanyi:2016ksw}. In  the improved EV-HRG model, we take the core radius 
$0.34$~fm  as a value of $r$, i.e., $b=0.63$~fm$^3$.   
For $\mu_{B}=400$~MeV  (lower panel), the EV-HRG result (solid line) agrees
with LQCD one~\cite{Borsanyi:2016ksw} in $T \lsim 210$~MeV, while the HRG result (dashed line) is consistent with LQCD one in $T \lsim 150$~MeV. 
For $\mu_{B}=0$~MeV (upper panel), both the EV-HRG and the HRG result are consistent with 
LQCD one~\cite{Borsanyi:2016ksw} in $ T \lsim 210$~MeV. 
The difference between the EV-HRG and HRG results means a repulsive
nature of  baryon and baryon.

\subsection{Independent quark model}
\label{Independent quark model}

We have to consider quark-gluon states in the region $T \gsim 200$~MeV. 
The Lagrangian density of the IQ model is 
\begin{eqnarray}
{\cal L}_{\rm Q}=\sum_f \left\{\bar{q}_f(i\gamma^{\mu}D_{\mu}-m_f )q_f \right\}-{\cal U}(T,{\Phi},{\bar{\Phi}}), 
\label{L_Q}
\end{eqnarray}
where $m_f$ is the current mass of $f$ quark 
and $D_{\mu}=\partial_{\mu}-igA_{\mu}^a\frac{\lambda_a}{2}\delta^{\mu0}$ with the Gell-Mann matrix $\lambda_a$ in color space. See  Refs.~\cite{Fukushima,Sakai} for the definition of the Polyakov loop ${\Phi}$ and its conjugate $\bar{\Phi}$. 

Making the path integral over quark fields leads to 
\begin{eqnarray}
P_{\rm Q}&=&\ -{\cal U}(T,{\Phi},{\bar{\Phi}}) 
\nonumber\\
&& + 2\sum_{f} \Bigg[ 
\int ( T\log{z_f^+}+ T\log{z_f^-})\Bigg] ,
\label{Omega_Q}
\end{eqnarray}
where 
\begin{eqnarray}
z_f^+ &=& 1+3{\bar{\Phi}} e^{-(E_f+\mu_f)/T}+3{\Phi} e^{-2(E_f+\mu_f)/T} 
\nonumber\\
&&+ e^{-3(E_f+\mu_f)/T},
\label{zfp}\\
z_f^- &=& 1+3{\Phi} e^{-(E_f-\mu_f)/T}+3{\bar{\Phi}} e^{-2(E_f-\mu_f)/T}\nonumber\\
&&+ e^{-3(E_f-\mu_f)/T}
\label{zfm}
\end{eqnarray}
with $E_f=\sqrt{{\bf p}^2+m_f^2}$. 
In Eq.~(\ref{Omega_Q}), the vacuum term has been omitted, since the pressure 
calculated with LQCD simulations does not include the term. 
The ${\Phi}$  and ${\bar{\Phi}}$ are obtained by minimizing   
$\Omega_{\rm Q}=-P_{\rm Q}$. 

The entropy density $s_{\rm Q}$ is obtained from 
$P_{\rm {\rm Q}}$ as
\begin{eqnarray}
s_{\rm Q}&=&{\partial P_{\rm Q}\over{\partial T}}. 
\label{s_Q}
\end{eqnarray}

We take the Polyakov-loop potential of Ref.~\cite{Miyahara:2017eam}:
\begin{eqnarray}
&&\frac{{\cal U}(T,{\Phi},{\bar{\Phi}})}{T^4}=-\frac{a(T)}{2}{\Phi}{\bar{\Phi}}\nonumber\\
&&+ b(T)\log\{1-6{\Phi}{\bar{\Phi}} + 4({\Phi}^3 + {\bar{\Phi}}^3) - 3({\Phi} {\bar{\Phi}})^2\};
\label{U_Phi}\\
&&a(T) = a_0 + a_1\left(\frac{T_0}{T}\right)+ a_2\left(\frac{T_0}{T}\right)^2,\\
&&b(T) = b_3\left(\frac{T_0}{T}\right)^3.   
\label{parameter_Phi}
\end{eqnarray}
The parameters $a_0$, $a_1$, $a_2$, $b_3$ and $T_0$ 
were fitted to 2+1 flavor $s_{\rm LQCD}$ in $400 \lsim T \lsim 500$~MeV; 
see Fig. 1 of Ref.~\cite{Miyahara:2017eam}  for the fit. 
The resulting values are tabulated in Table~\ref{Table_cal_U}.

\begin{table}[h]
\centering
\caption{Parameters in the Polyakov-loop potential. 
}
\begin{tabular}{lcccr}
\hline
$\ a_0$&$a_1$&$a_2$&$b_3$&$T_0\hspace{5mm}$
\\ \hline
2.457 &-2.47&15.2&-1.75&270{\rm [MeV]}\\
\hline
\end{tabular}
\label{Table_cal_U}
\end{table}

\subsection{sHQH model}
\label{improved Quark-hadron hybrid model}


The total entropy reads 
\begin{eqnarray}
s(T,\mu_B)&=&f_{\rm H}(T) s_{\rm inv:H}(T,\mu_B) 
\nonumber \\
&~& + \left[1-f_{\rm H}(T) \right] 
s_{\rm Q}(T,\mu_B)   
\label{s_hybrid_def-2}
\end{eqnarray}
in the sHQH model. 
We consider that the $f_{\rm H}(T,\mu_B)$ has no $\mu_B$ dependence, 
since $s_{\rm inv:H}$ and $s_{\rm Q}$ depend on  $\mu_B$.
This allows us to determine $f_{\rm H}(T)$ from $s=s_{\rm LQCD}$~\cite{Borsanyi:2016ksw}  
at $\mu_B=0$: 
Namely,  
\bea
f_{\rm H}(T)=\frac{s_{\rm LQCD}(T)-s_{\rm Q}(T)}{s_{\rm inv:H}(T)-s_{\rm Q}(T)} .  
\label{fH-LQCD}
\eea

\begin{figure}[h]
\centering
\vspace{0cm}
\includegraphics[width=0.4\textwidth]{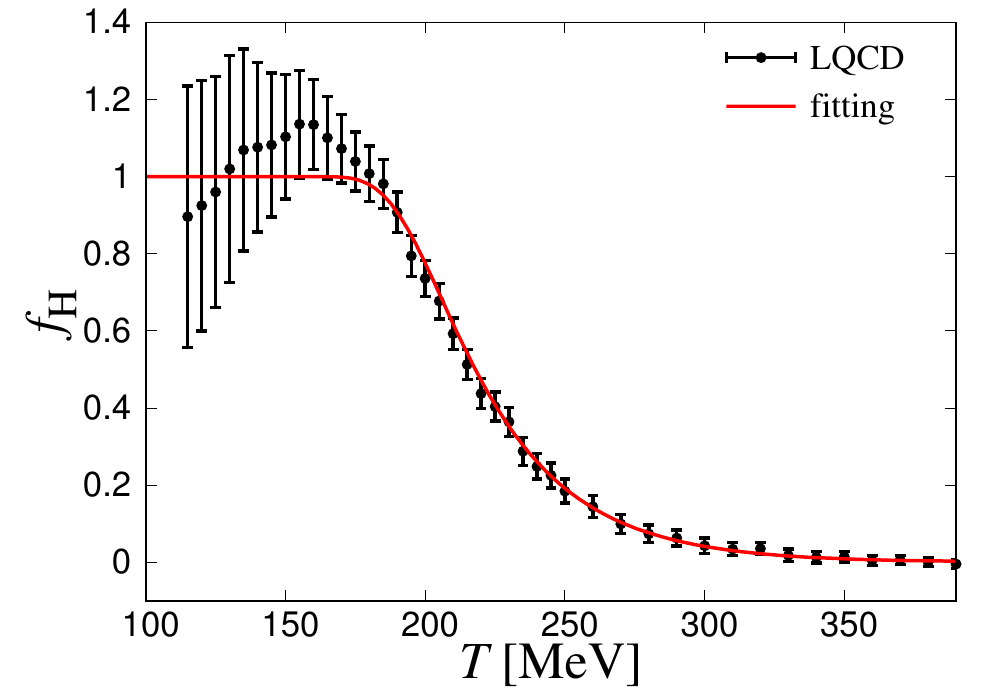}
\vspace{-10pt}
\caption{$T$ dependence of the switching function $f_{\rm H}(T)$. 
The dots with error bars are the  $f_{\rm H}(T)$ of Eq.~\eqref{fH-LQCD}, 
The solid line is a fitting function for  
the  $f_{\rm H}(T)$; see the text for the fitting. 
}
\label{fig_f-H}
\end{figure}

In Fig.~\ref{fig_f-H}, the  $f_{\rm H}(T)$ of Eq.~\eqref{fH-LQCD} is shown 
by dots with error bars. The errors come from $s_{\rm LQCD}$. 
The solid line is a fitting function for 
the  $f_{\rm H}(T)$ of Eq.~\eqref{fH-LQCD}; in the $\chi^2$ fitting, the line 
is assumed to be 1 in $T<180$~MeV. 
From now on, we regard the solid line as  the switching function $f_{\rm H}(T)$. 
The weight function $f_{\rm H}(T)$ means the occupancy of
hadronic matter in the total entropy, and the condition $0\le f_{\rm H}\le 1$
should be satisfied.

The pressure $P$ with no vacuum contribution is obtainable from 
$s_{\rm LQCD}$ of Eq.~\eqref{s_hybrid_def-2}:
\begin{eqnarray}
P(T,\mu_{B})
=\int_{0}^{T}dT' s(T',\mu_{B})
\label{P_hybrid-2}
\end{eqnarray}
The energy density is obtained by 
$
\varepsilon(T,\mu_B)=sT-P+\mu_B n, 
$
where $n$ is the baryon-number density.

\vspace{-0pt}
\section{Numerical results}
\label{results}
\vspace{-0pt}

As mentioned in Sec.~\ref{Introduction}, 
we can consider the transition region determined from  with 
the peak and the half-value width of $d \varepsilon(T,\mu_{B})/d T$ as 
a chiral-transition region. This statement is confirmed explicitly by analyses shown in this section.

\subsection{$T$ dependence of the Polyakov loop for $\mu_{B}=0 \sim 400$~MeV}

\begin{figure}[h]
\centering
\vspace{0cm}
\includegraphics[width=0.4\textwidth]{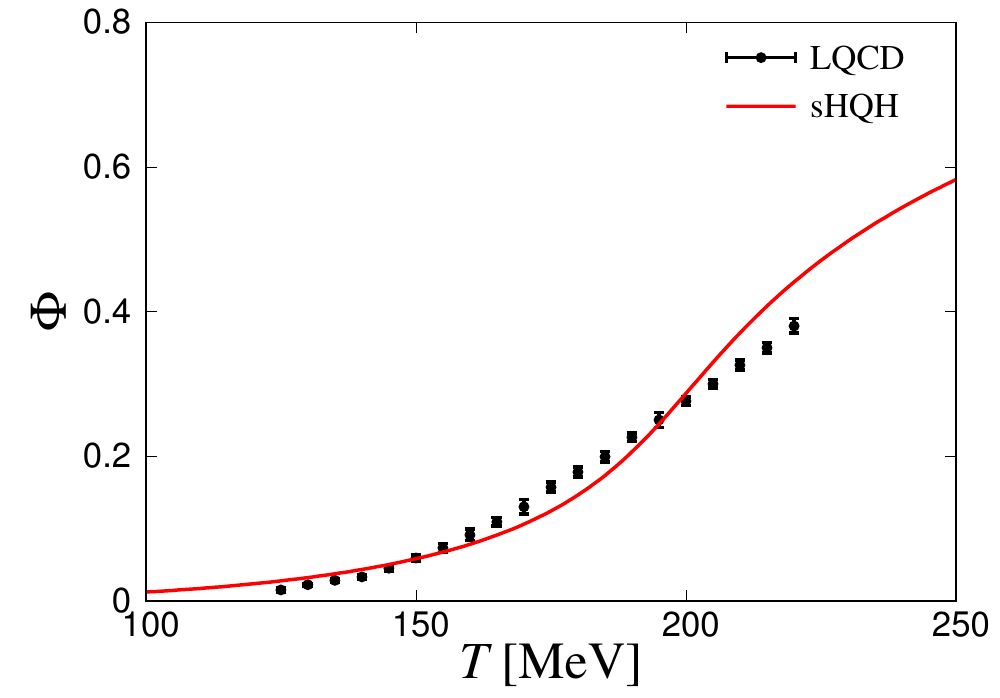}
\includegraphics[width=0.4\textwidth]{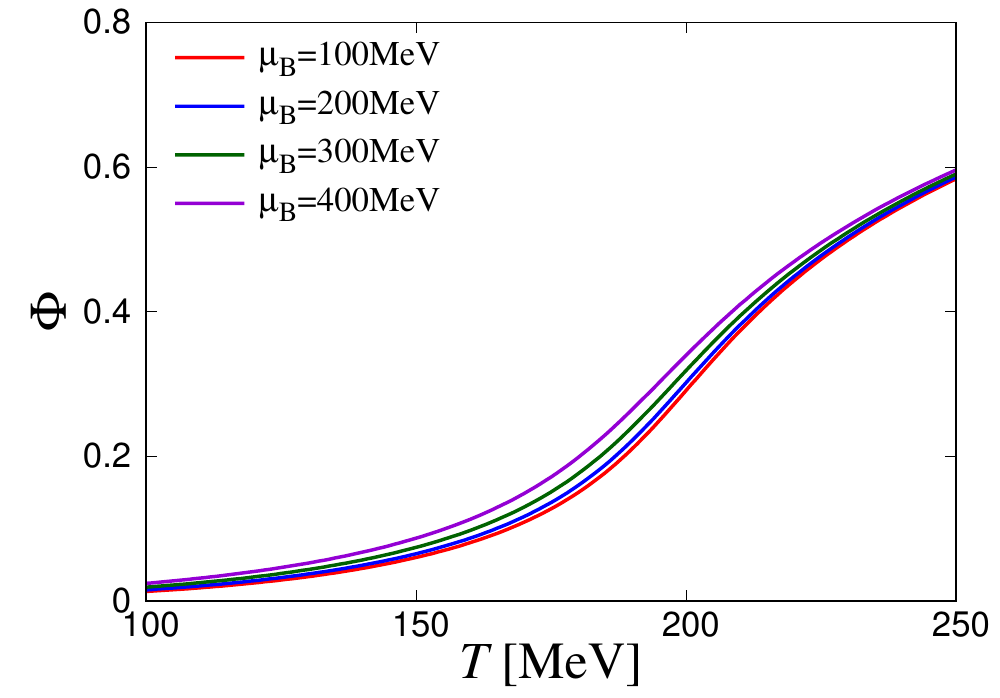}
\vspace{-10pt}
\caption{$T$ dependence of the Polyakov loop ${\Phi}$. 
The upper panel is for $\mu_{B}=0$~MeV and the lower panel is 
for $\mu_{B}=100, 200, 300, 400$~MeV. 
The sHQH model results are shown by the solid lines. 
In the lower panel, four lines correspond to the cases of 
$\mu_{B}=100, 200, 300, 400$~MeV from right to left. 
LQCD data are taken from 
Ref.~\cite{Borsanyi:2010bp}. 
}
\label{fig:$T$ dependence of the Polyakov loop}
\end{figure}

Figure~\ref{fig:$T$ dependence of the Polyakov loop} shows the Polyakov loop 
${\Phi}$ as a function of $T$ for the cases of 
$\mu_{B}=0, 100, 200, 300,400$~MeV. 
The LQCD result is available only for $\mu_{B}=0$~MeV~\cite{Borsanyi:2010bp}. 
In the upper panel for $\mu_{B}=0$~MeV, the sHQH  result (solid line) 
well reproduces LQCD one in which the continuum limit is taken.   
We then predict the ${\Phi}$ for $\mu_{B}=100, 200, 300, 400$~MeV in the 
lower panel. $\mu_{B}$ dependence of ${\Phi}$ is small.

\subsection{Transitions}
\label{QCD Transition}

We first consider the case of  $\mu_{B}=0$.
Table~\ref{Tc_summary} shows  results of sHQH model for  the transition region $T_{\rm c}^{\varepsilon}$ determined 
from the peak  and the half-valued width of $d \varepsilon(T,\mu_{B})/d T$.  The result is compared with 
LQCD data~\cite{Borsanyi:2010bp} on the chiral transition temperature $T_{\rm c}^{\chi:LQCD}$. 
Obviously, $T_{\rm c}^{\Delta_{l,s}:LQCD}$ is in the region $T_{\rm c}^{e}$. 

\begin{table}[h]
\centering
\begin{tabular}{lccr}
\hline
$\hspace{5mm}T_{\rm c}^{\varepsilon}$&$T_{\rm c}^{\Delta_{l,s}:LQCD}
\hspace{5mm}$\\ \hline
137--204{\rm [MeV]}&157(4)(3){\rm [MeV]}
\\ \hline
\end{tabular}
\caption{Comparison between lattice transition temperature and transition region calculated with sHQH model 
for $\mu_{B}=0$. 
}
\label{Tc_summary}
\end{table}

\begin{figure}[h]
\centering
\vspace{0cm}
\includegraphics[width=0.4\textwidth]{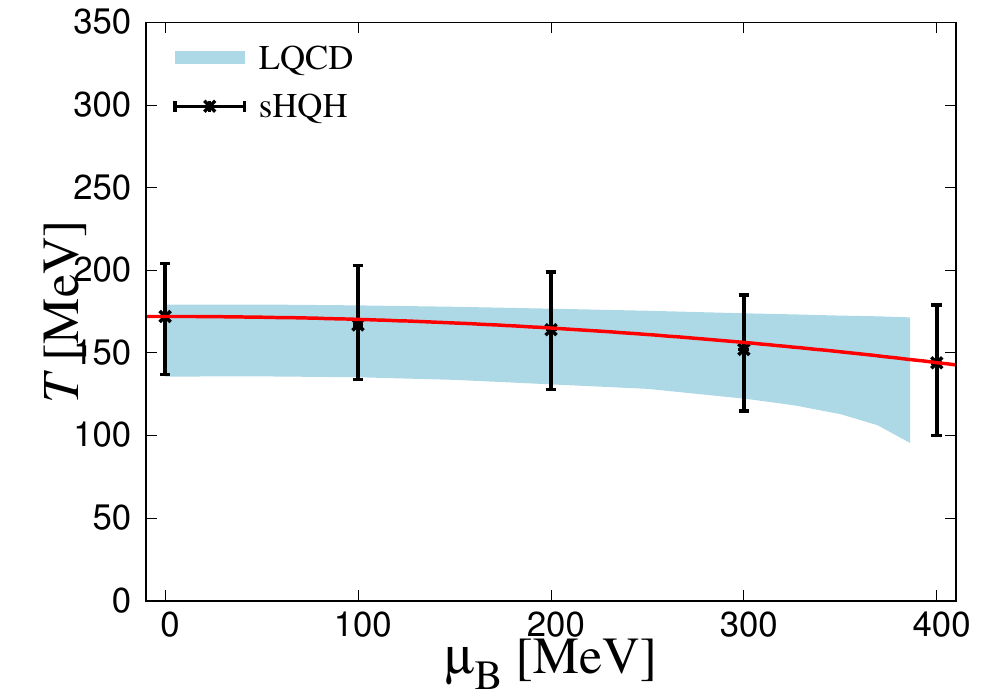}
\vspace{-10pt}
\caption{Crossover region determined from $d \varepsilon/dT$ in $\mu_{B}$--$T$ plane. 
The blue band is the chiral-transition region determined by 
analytic continuation of LQCD simulations from imaginary to real $\mu$~\cite{Bellwied:2015rza}. 
The horizontal line with cross stands for the transition region determined from $d \varepsilon/dT$ and 
is  calculated with the sHQH model. The transition line  (red solid line), obtained by connecting the crosses,  
is expressed by $T = 172(1 - 0.038(\mu_{B}/172)^2)$~MeV. 
}
\label{fig_chiral-transition-line}
\end{figure}

Figure~\ref{fig_chiral-transition-line} shows  the transition region determined 
from the peak  and the half-valued width of $d \varepsilon(T,\mu_{B})/d T$ and the lattice chiral-transition region 
in $\mu_{B}$--$T$ plane; the former is  calculated with the sHQH model and the latter is analytic 
continuation of LQCD  simulations from imaginary to real $\mu$~\cite{Bellwied:2015rza}.  
The transition region determined 
from $d \varepsilon(T,\mu_{B})/d T$ is shown  by a horizontal line with cross 
for each of $\mu_{B}=0, 100, 200, 300, 400$~MeV; 
the cross is a maximum value of  $d \varepsilon/dT$ and the horizontal line means 
the half-value width of $d \varepsilon/dT$. 
The red solid line is made by connecting the crosses. 
Meanwhile, the blue band indicates the width 
of the chiral-transition region extrapolated from the imaginary-$\mu_B$ 
region~\cite{Bellwied:2015rza}. 
The transition region calculated with the sHQH model is consistent with the LQCD result. 
We can thus regard the transition region determined from $d \varepsilon(T,\mu_{B})/d T$ as a chiral-crossover 
region.

As shown in the right panel of Fig. 4 of Ref.~\cite{Nonaka:2004pg}, Nonaka and Asakawa showed 
that in $\mu_{B}$--$T$ plane the isentropic trajectories are focused to the CEP. 
They concluded that the CEP acts as an attractor of isentropic trajectories, 
 $n/s=$const.

\begin{figure}[h]
\centering
\vspace{0cm}
\includegraphics[width=0.4\textwidth]{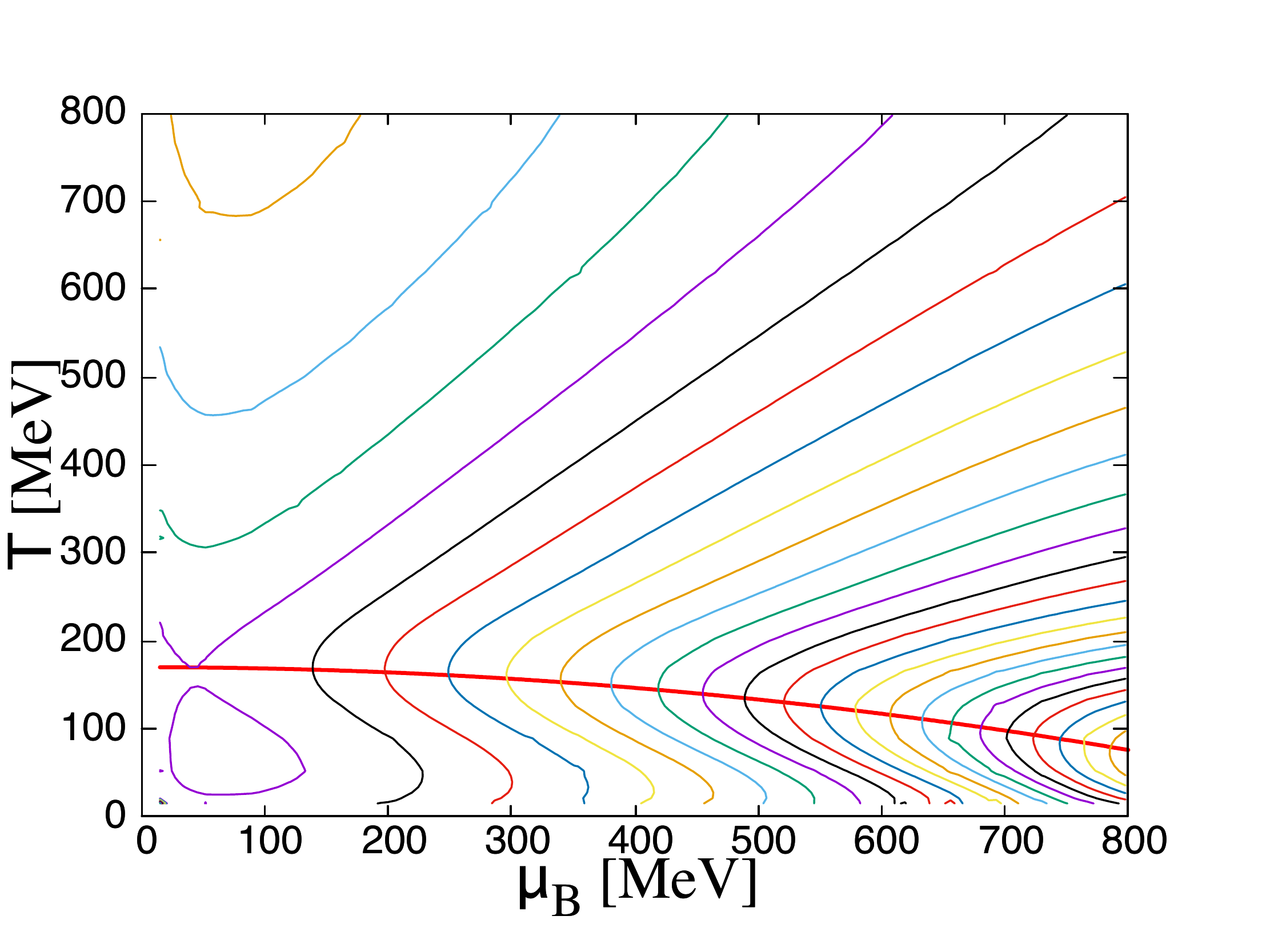}
\includegraphics[width=0.4\textwidth]{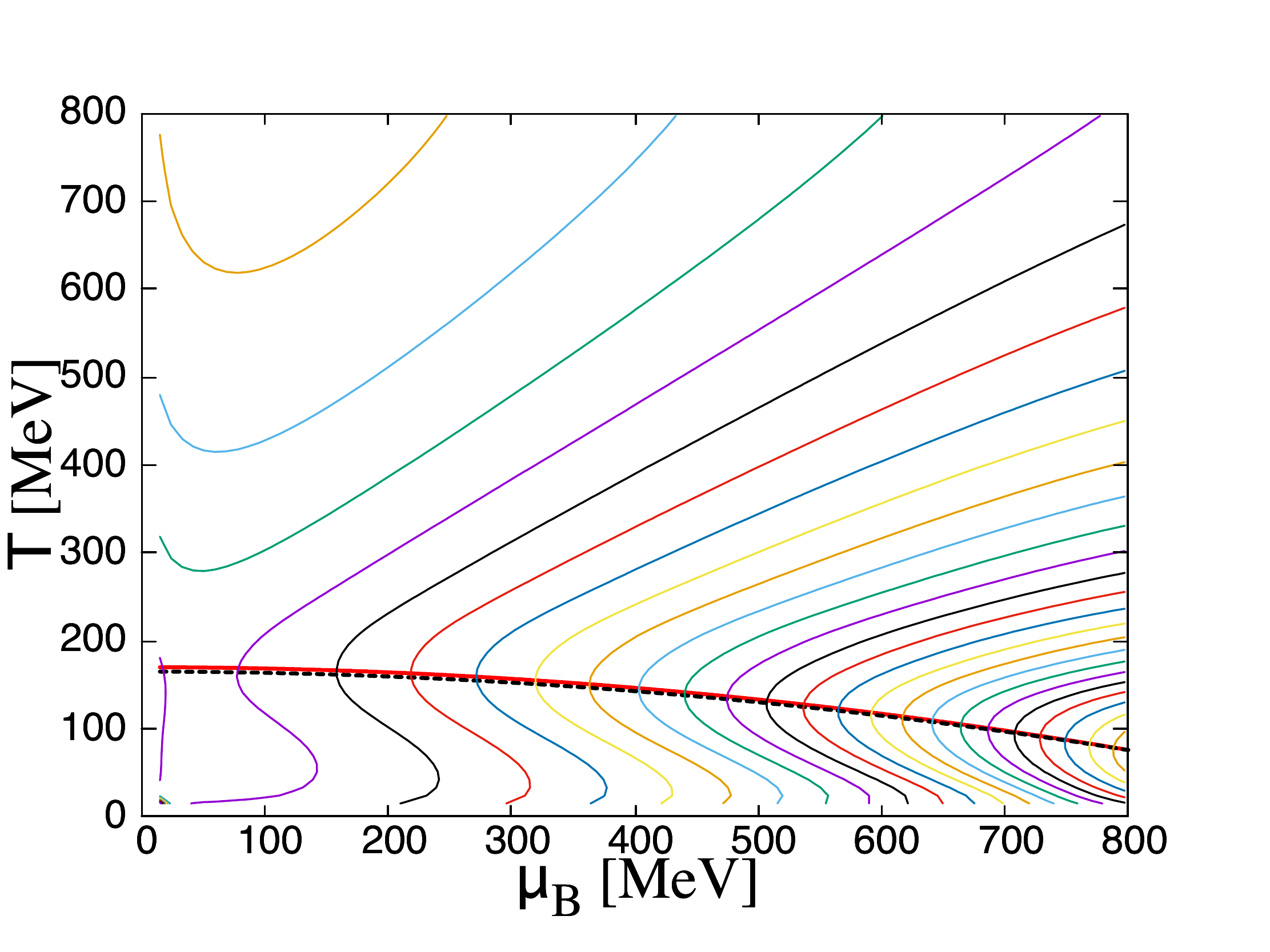}
\vspace{-10pt}
\caption{
Isentropic trajectories, $n/s$=const, in $\mu_{B}$--$T$ plane where  
the strangeness neutrality is imposed for the lower panel and not for the upper panel.  
In the upper panel, the solid curve is a line connecting the points 
at which the curve of trajectory becomes maximum; 
the resulting curve is $T = 170(1 - 0.025(\mu_{B}/170)^2)$~MeV. 
The isentropic trajectories are shown by $n/s=0.014 \sim 0.072$ 
from top left to bottom right. 
In the lower panel, the dashed line stands for a  transition line with the strangeness neutrality, i.e., 
 $T = 165(1 - 0.023(\mu_{B}/165)^2)$~MeV. For comparison, we also show the the solid line 
 $T = 170(1 - 0.025(\mu_{B}/170)^2)$~MeV  in which the strangeness neutrality is not imposed. 
The isentropic trajectories are shown by $n/s=0.012 \sim 0.07$ 
from top left to bottom right. 
}
\label{fig_Isentropic-trajectories}
\end{figure}

In the upper panel of Fig.~\ref{fig_Isentropic-trajectories}, 
the solid curve is a line connecting the points 
at which the curvature of isentropic trajectory becomes maximum. 
The curve is connected to  the CEP, if it exists~\cite{Nonaka:2004pg}. 
We can thus regard the curve as a transition line in $\mu_{B}$--$T$ plane. 

In the lower panel, we impose  the strangeness neutrality. 
Comparing the two panels, we can find that the effect of  strangeness neutrality is small. 
Hence, the transition calculated with $n/s$ may be 
deduced from relativistic nuclear collisions. There is no evidence 
of focusing (attractor) of isentropic trajectory in the sHQH model.

\begin{figure}[H]
\centering
\vspace{0cm}
\includegraphics[width=0.4\textwidth]{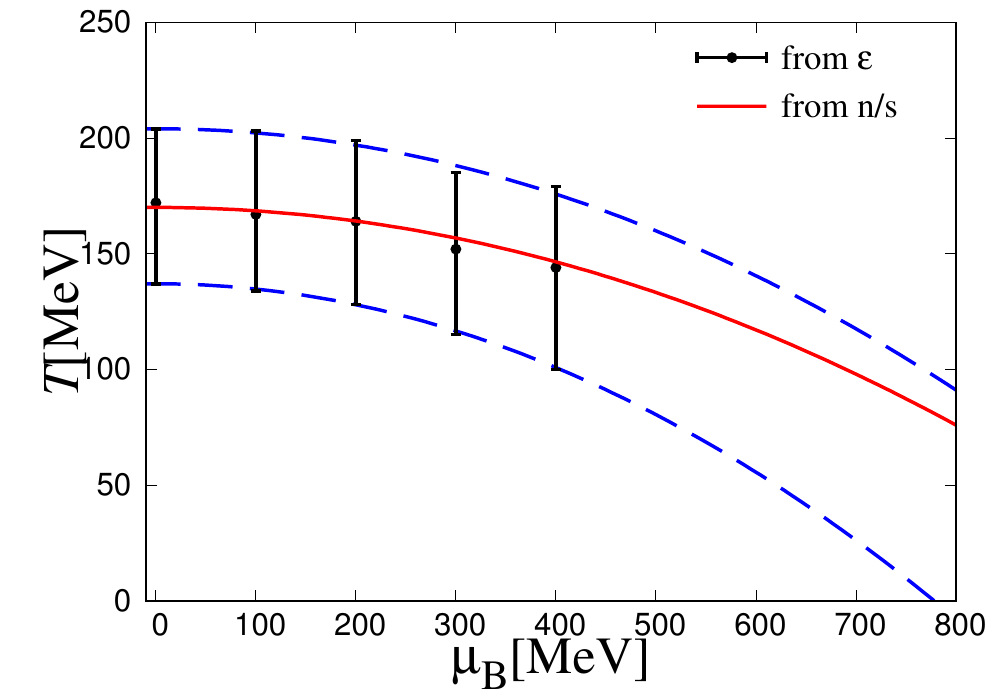}
\vspace{-10pt}
\caption{
Chiral-crossover region  determined from $d \varepsilon/dT$ and transition line  determined from $n/s$ 
in $\mu_{B}$--$T$ plane.  
The chiral-crossover region determined from the peak and the half-valued width of $d \varepsilon/dT$
is denoted for $\mu_{B}=0, 100, 200, 300, 400$~MeV by dots with errorbars. The upper and lower sides 
of chiral-crossover region are shown by two dashed lines. 
The transition line determined from $n/s$ is $T = 170(1 - 0.025(\mu_{B}/170)^2)$~MeV. 
}
\label{fig_transaction-lines}
\end{figure}

Figure~\ref{fig_transaction-lines} shows the transition line determined from $s/n$  by a solid line 
and  the chiral-crossover region from the peak and the half-valued width of $d \varepsilon/dT$ 
by two dashed lines in $\mu_{B}$--$T$ plane. 
Here we do not consider the strangeness neutrality, because the effect is small. 
The transition line obtained from $s/n$  passes in the vicinity of dots (the peak of $d \varepsilon/dT$) 
and is in the chiral-crossover region. 
This allows us to regard the transition line etermined from $s/n$ as a chiral-transition line. 
The quantity   $s/n$ is quite useful, since it is obtainable from not ony LQCD but also heavy-ion collisions.

\subsection{The EoS}

In this section, we do not consider the strangeness neutrality, because the effect is small. 

In order to compare the present model with the previous model~\cite{Miyahara:2017eam}, we take the same assumption 
`` $f_{\rm H}(T)$ has no $\mu_{B}$ dependence'', in the the previous model. 
The resulting switching function ${f}_{\rm H}^{\rm prev}(T)$ is shifted 
to the left by about 10~MeV from $f_{\rm H}(T)$ in Fig.~\ref{fig_f-H}. 
The difference between the present model with $f_{\rm H}(T)$ and the previous model with ${f}_{\rm H}^{\rm prev}(T)$ shows EV effects. 
The previous model with ${f}_{\rm H}^{\rm prev}(T)$ is referred to as 
``HRG-HQH model'' in this paper.

\begin{figure}[H]
\centering
\vspace{0cm}
\includegraphics[width=0.4\textwidth]{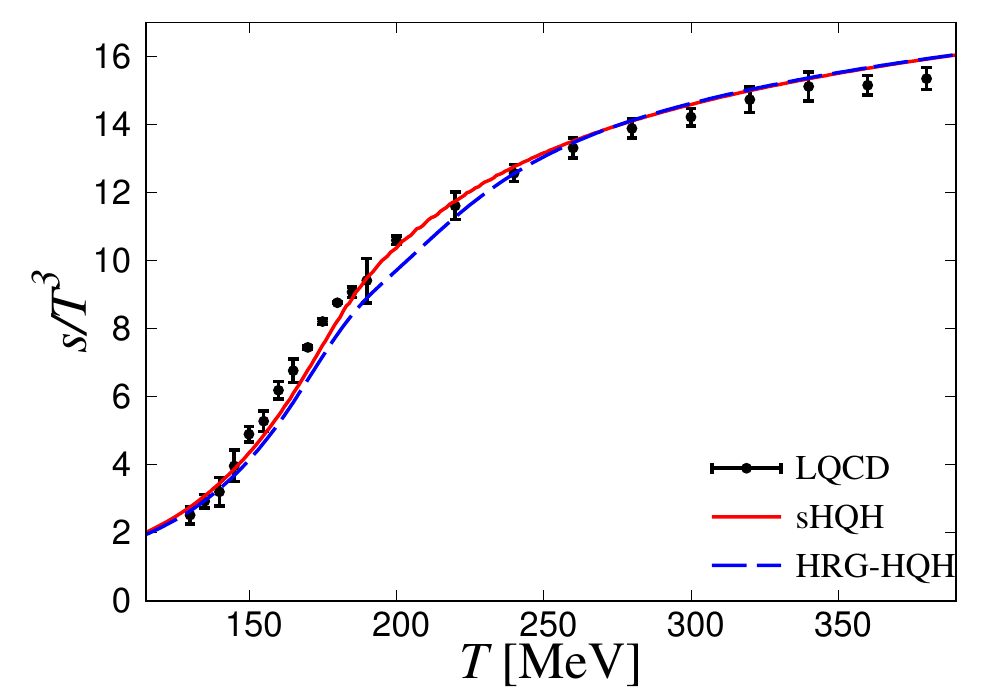}
\includegraphics[width=0.4\textwidth]{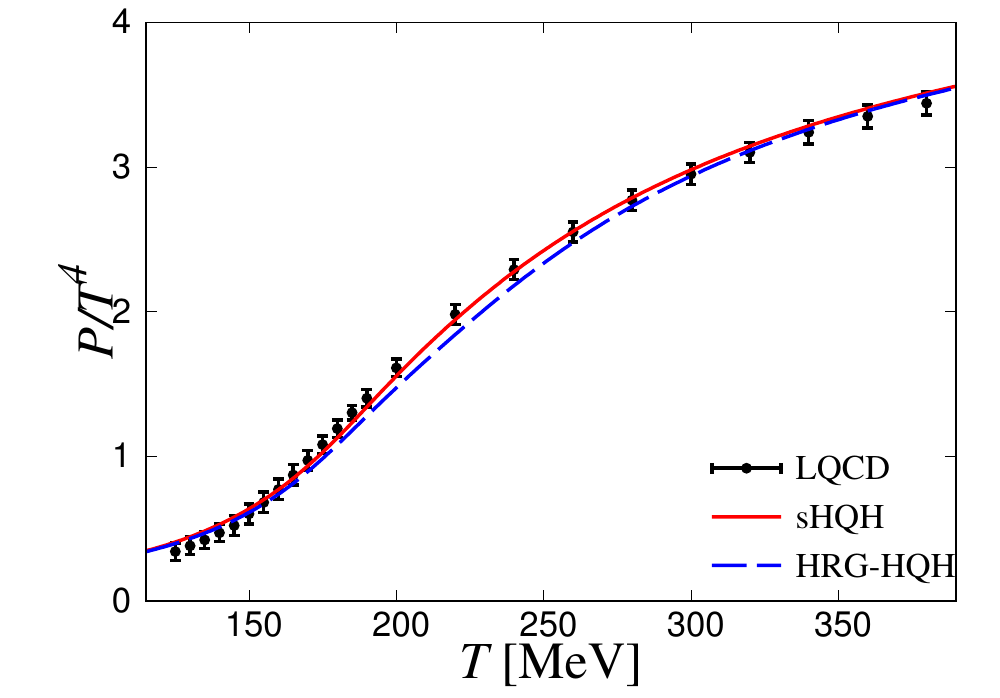}
\includegraphics[width=0.4\textwidth]{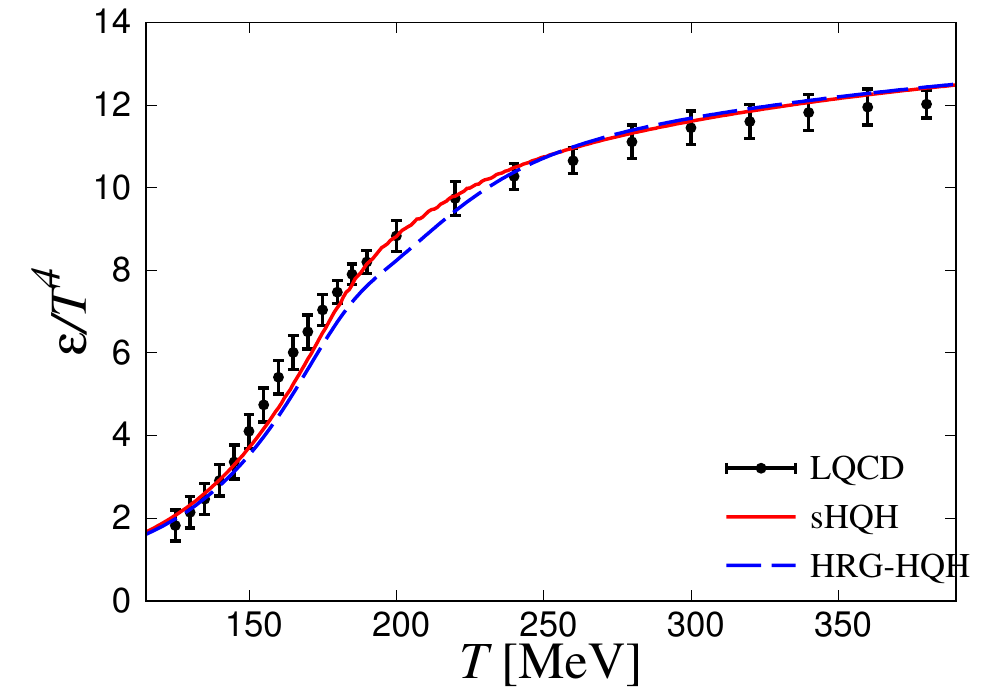}
\vspace{-10pt}
\caption{$T$ dependence of $s$, $P$, $\varepsilon$ 
at $\mu_{B}=0$~MeV. 
See the the text for the definition of lines. 
LQCD data are taken from Ref.~\cite{Borsanyi:2012cr}.
}
\label{fig_Eos-0}
\end{figure}

Figure~\ref{fig_Eos-0} shows $T$ dependence of $s$, $P$, $\varepsilon$, 
at $\mu_{B}=0$~MeV. 
The solid and  dashed lines are the results of sHQH and HRG-HQH models, 
respectively. The difference between the two lines shows EV effects. 
we can find that the effects are small for $\mu_{B}=0$~MeV.
We find that the fitting of $f_H(T)$ is good, since the sHQH result 
agrees with LQCD data~\cite{Borsanyi:2012cr}. Also for $P$  and $\varepsilon$, 
the sHQH  model reproduces LQCD data.

\begin{figure}[H]
\centering
\vspace{0cm}
\includegraphics[width=0.4\textwidth]{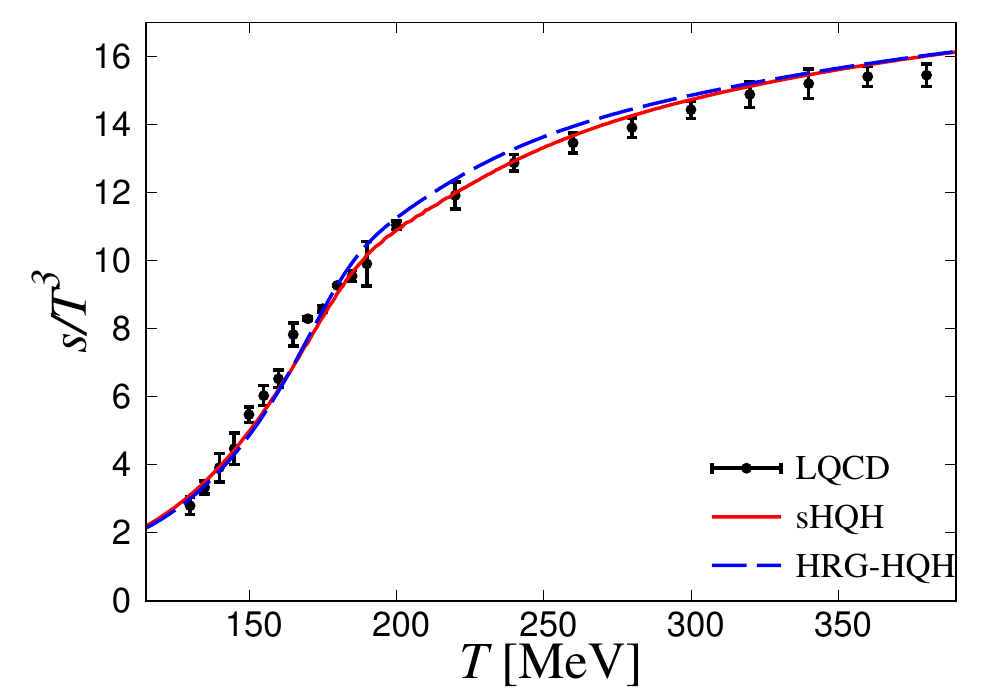}
\includegraphics[width=0.4\textwidth]{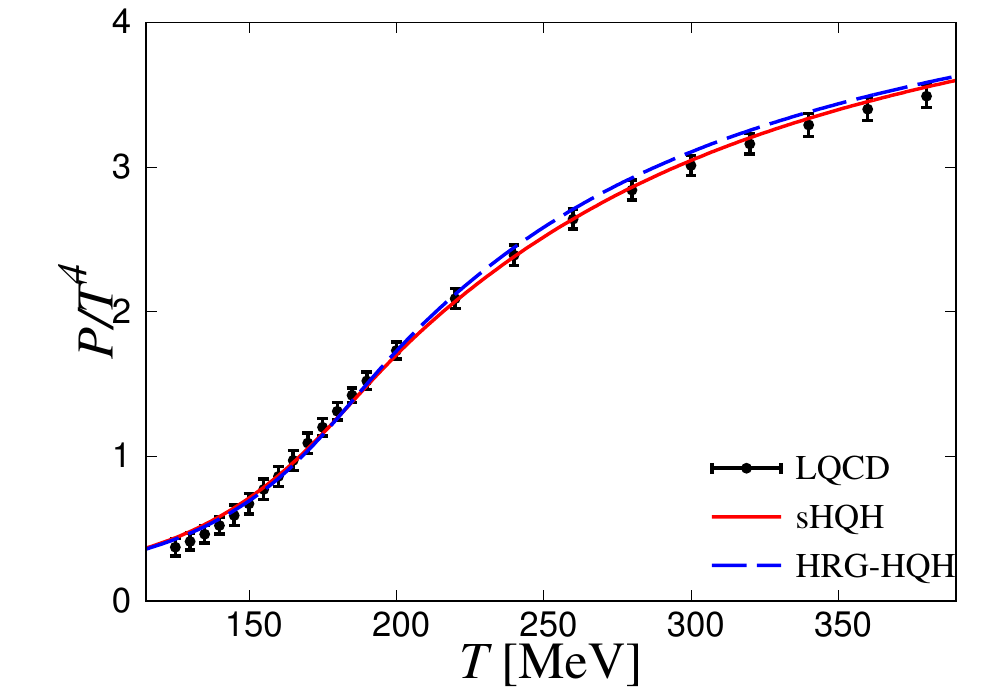}
\includegraphics[width=0.4\textwidth]{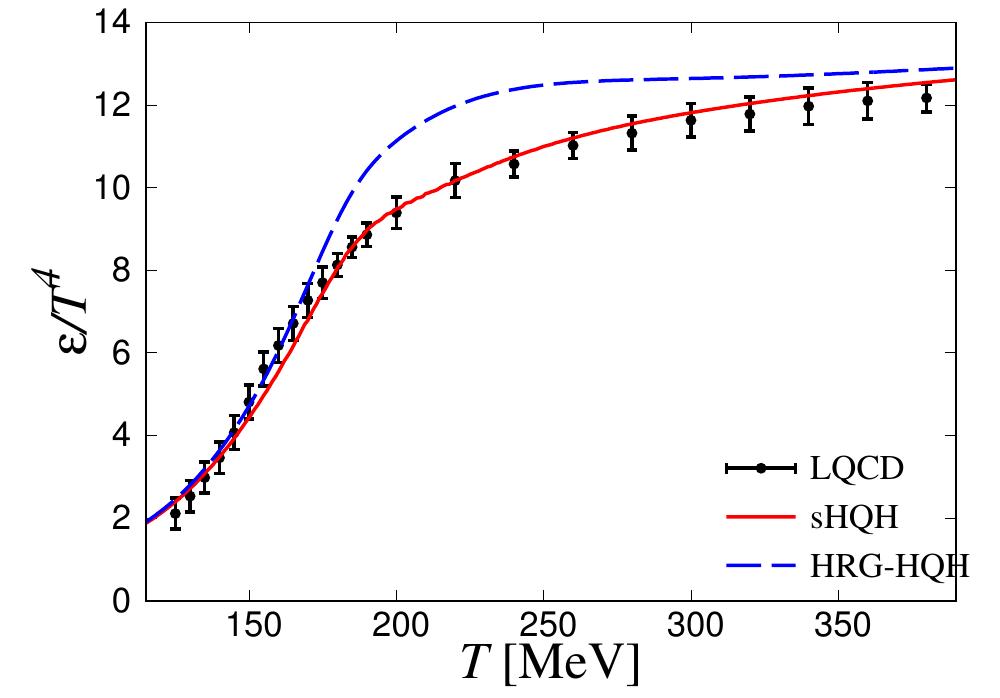}
\includegraphics[width=0.4\textwidth]{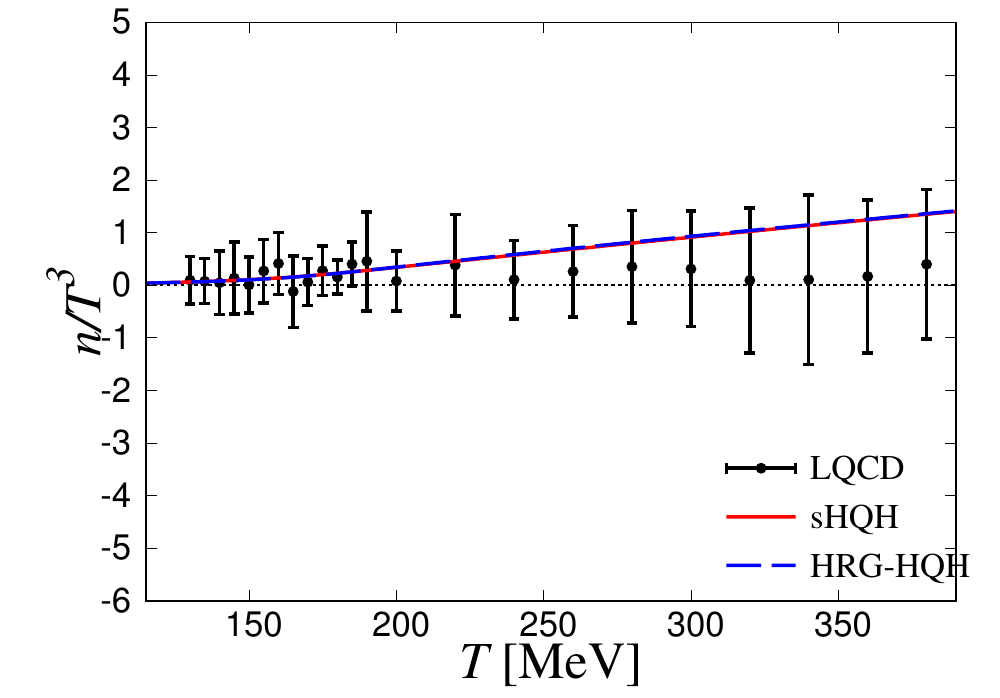}
\vspace{-10pt}
\caption{
$T$ dependence of $s$, $P$, $\varepsilon$, $n$ 
at $\mu_{B}=200$~MeV. 
See the the text for the definition of lines. 
LQCD data are taken from Ref.~\cite{Borsanyi:2012cr}; 
note that $n$ is deduced from $s$, $P$, $\varepsilon$. 
}
\label{fig_Eos-2}
\end{figure}

\begin{figure}[H]
\centering
\vspace{0cm}
\includegraphics[width=0.4\textwidth]{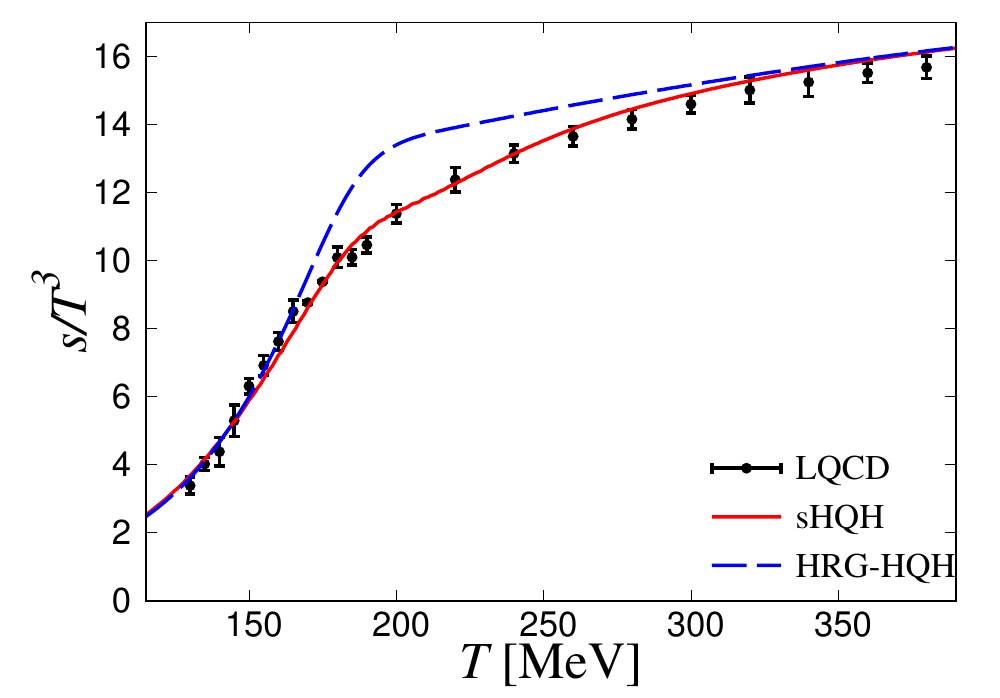}
\includegraphics[width=0.4\textwidth]{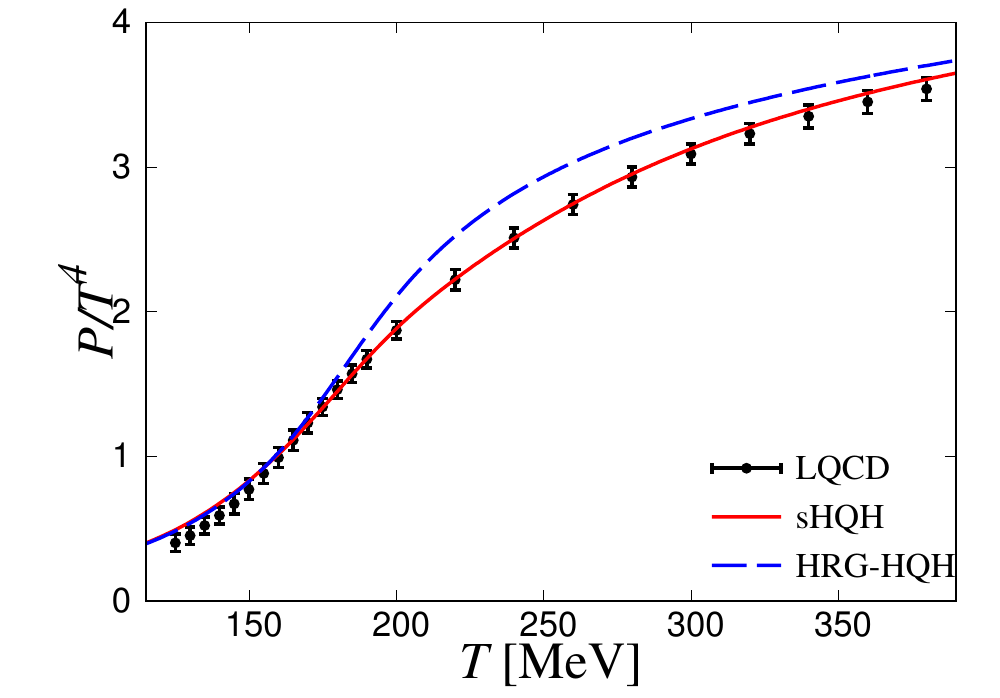}
\includegraphics[width=0.4\textwidth]{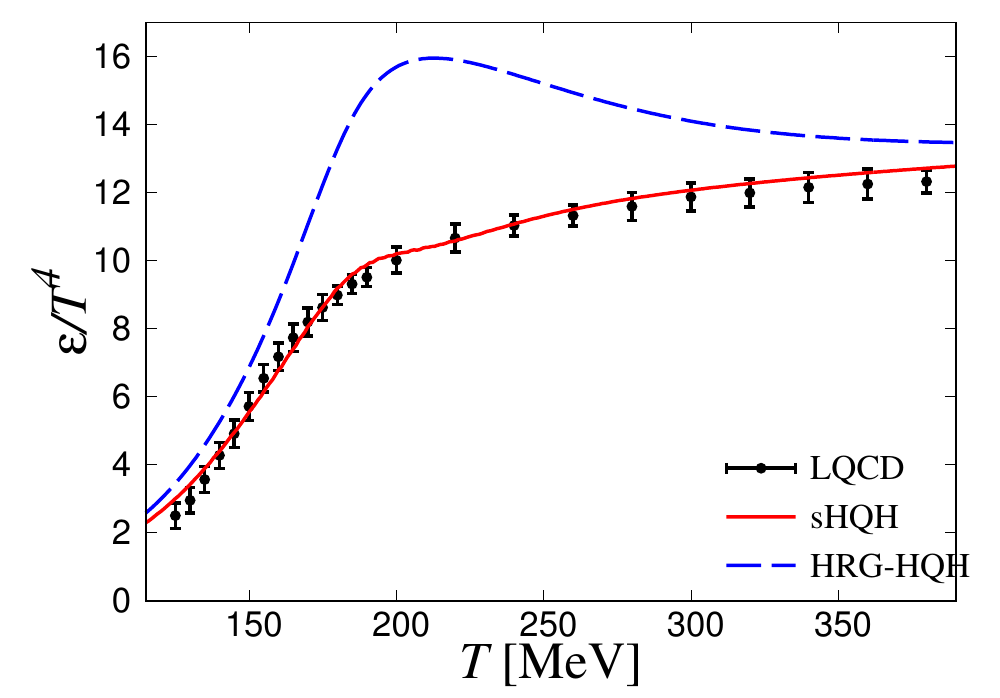}
\includegraphics[width=0.4\textwidth]{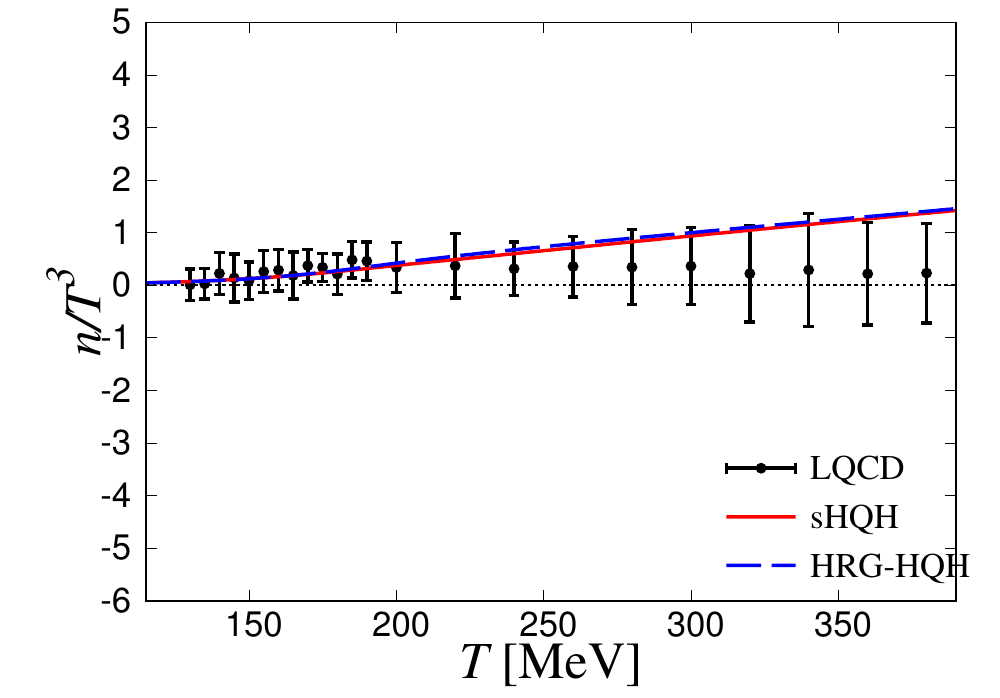}
\vspace{-10pt}
\caption{
$T$ dependence of $s$, $P$, $\varepsilon$, $n$ 
at $\mu_{B}=300$~MeV. 
See the the text for the definition of lines. 
LQCD are taken from Ref.~\cite{Borsanyi:2012cr}; 
note that $n$ is deduced from $s$, $P$, $\varepsilon$.  
}
\label{fig_Eos-3}
\end{figure}

\begin{figure}[H]
\centering
\vspace{0cm}
\includegraphics[width=0.4\textwidth]{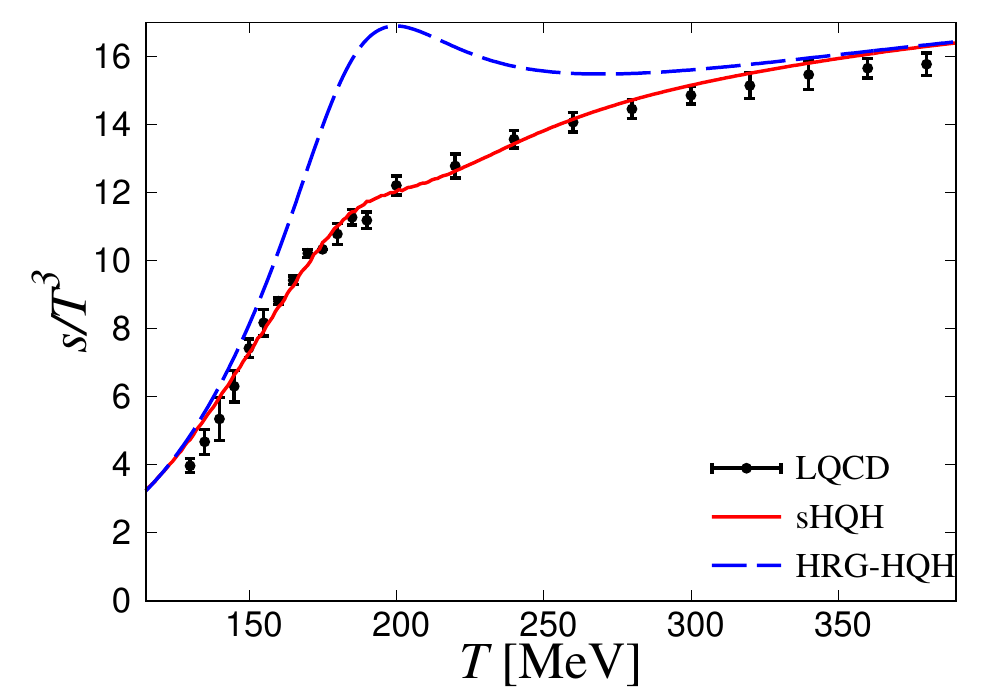}
\includegraphics[width=0.4\textwidth]{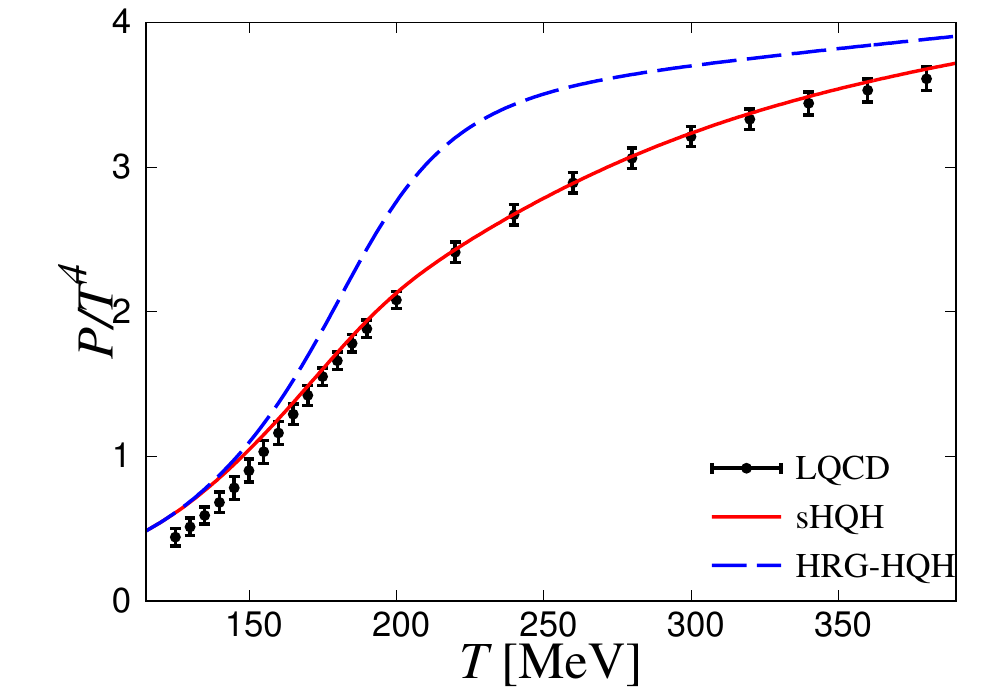}
\includegraphics[width=0.4\textwidth]{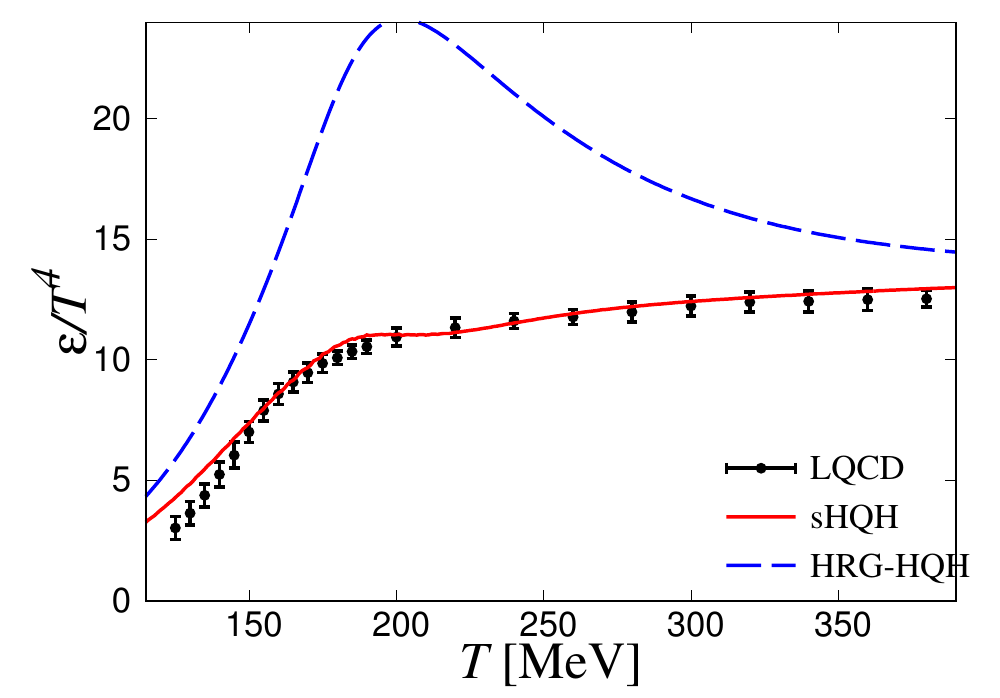}
\includegraphics[width=0.4\textwidth]{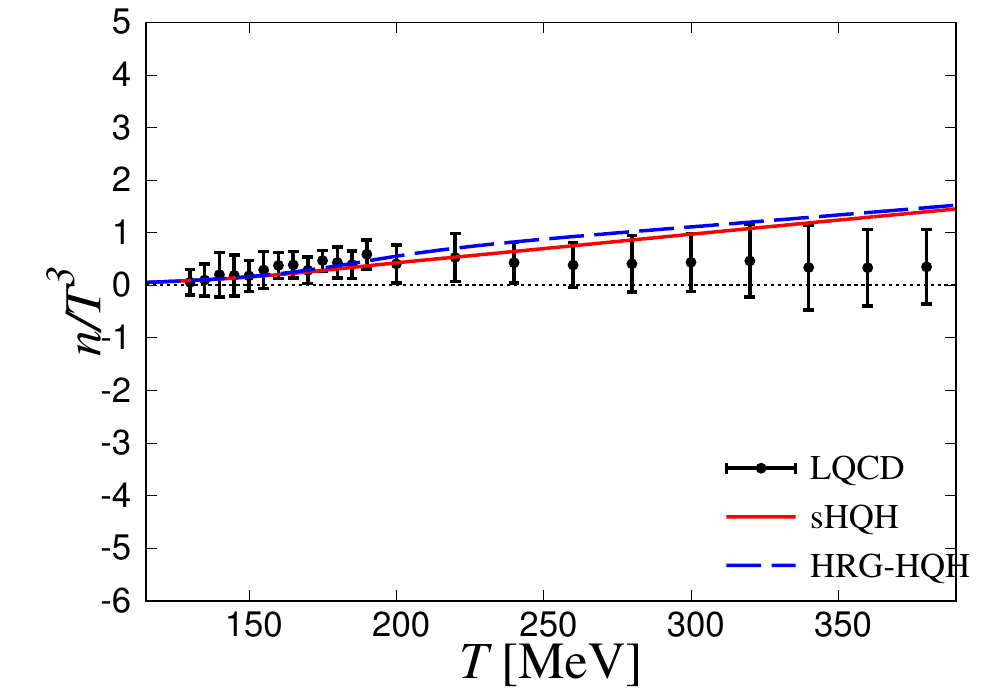}
\vspace{-10pt}
\caption{
$T$ dependence of $s$, $P$, $\varepsilon$, $n$ 
at $\mu_{B}=400$~MeV. 
See the the text for the definition of lines. 
LQCD are taken from Ref.~\cite{Borsanyi:2012cr}; 
note that $n$ is deduced from $s$, $P$, $\varepsilon$.  
}
\label{fig_Eos-4}
\end{figure}

Figures~\ref{fig_Eos-2}--\ref{fig_Eos-4} shows $T$ dependence of $s$, $P$, $\varepsilon$, $n$ 
for $\mu_{B}=200, 300, 400$~MeV. 
The results of sHQH model well reproduces  the LQCD data~\cite{Borsanyi:2012cr}. 
 EV effects become large as $\mu_{B}$ increases from 200~MeV to 400~MeV. 
 This means that the interaction between baryons becomes non-negligible as $\mu_{B}$ increases.

\section{Summary}
\label{summary}

We have constructed a simple HQH (sHGH) model in the $\mu_B$--$T$ plain, 
improving the EV-HRG model~\cite{Vovchenko:2014pka,Vovchenko:2015cbk} 
for the hadron piece and using the simple IQ model for the quark-gluon piece. 
The improved EV-HRG model yields the baryon and antibaryon pressures 
as simple analytic functions of Eqs.~\eqref{EQ:PB_mod-ell=1}--\eqref{EQ:PaB_mod-ell=1}, 
and ensures that the pressure is $\mu_B$ even.

As an interesting result of LQCD simulations for $\mu_B=0$~\cite{Borsanyi:2010bp}, 
the chiral-crossover region determined from $d \Delta_{\rm l,s}/dT$
agrees with the region from $d \varepsilon(T,\mu_B)/dT$. 
In LQCD simulations for finite $\mu_B$~\cite{Borsanyi:2012cr}, furthermore, a 
transition region is obtained by $d \varepsilon(T,\mu_B)/dT$. 
We may regard the transition region determined from  $\varepsilon$ as 
a chiral-crossover region. 
In fact, the crossover region  determined from $d \varepsilon(T,\mu_B)/dT$ of sHQH model 
agrees with the 
lattice result for the chiral-crossover region~\cite{Bellwied:2015rza} in $\mu_B \leq 400$~MeV. 
We have then predicted the chiral-crossover region in $ 400 \leq \mu_B \leq 800$~MeV.

In this work, we have considered that the $f_{\rm H}(T,\mu_B)$ does not depend on $\mu_B$, 
since $s_{\rm inv:H}$ and $s_{\rm Q}$ depend on  $\mu_B$.
This allows us to determine the switching function $f_{\rm H}(T)$  from 
$s_{\rm LQCD}$ at $\mu_B=0$. 
The present sHQH with the $f_{\rm H}(T)$ is successful in 
reproducing LQCD data on not only the chiral transition region but also the EoS in $\mu_{B} \leq 400$~MeV. 
In addition, the present sHQH model  accounts for 
LQCD data on the Polyakov loop at $\mu_{B} =0$~MeV. We have then predicted  
the Polyakov loop for $\mu_{B} =100, 200, 300, 400$~MeV.

Using the simple-HQH model, we have also predicted  a transition line derived from isentropic trajectories
in $0 \leq \mu_B \leq 800$~MeV. We found that there is no evidence of attractor of isentropic trajectories 
and the effect of strange neutrality is small for the transition line derived from isentropic trajectories. 
Further analyses of these properties seem to be important for both LQCD and relativistic nuclear collisions.

\vspace{-0pt}
\noindent
\begin{acknowledgments}
\vspace{-0pt}
The authors thank Junpei Sugano and Takehiro  Hirakida for useful 
contributions. 
H. K. is supported 
by Grant-in-Aid for Scientific Research (No.17K05446) 
from the Japan Society for the Promotion of Science (JSPS). 
\end{acknowledgments}

\appendix


\end{document}